\documentclass[prd,superscriptaddress,unsortedaddress,twocolumn,showpacs,floatfix]{revtex4}

\usepackage{epsfig}
\usepackage{dcolumn}
\usepackage{amsmath}

\newcommand{\vusdef}{\mbox{$\Gamma_{K\ell 3}=\frac{\textstyle G_F^2
M_K^5}{\textstyle 192\pi^3} S_{EW} (1+\delta^\ell_K)
\left|\vus\right|^2|f^2_+(0)| I^\ell_K   $}}
\newcommand{\ppzkin}{\mbox{$k_{+-0} $}}

\newcommand{\lplus}{\mbox{$\lambda_+$}}
\newcommand{\lzero}{\mbox{$\lambda_0$}}
\newcommand{\lplusp}{\mbox{$\lambda'_+$}}
\newcommand{\lzerop}{\mbox{$\lambda'_0$}}
\newcommand{\lpluspp}{\mbox{$\lambda''_+$}}
\newcommand{\lzeropp}{\mbox{$\lambda''_0$}}

\newcommand{\mpilep}{M_{\pi\ell}}
\newcommand{\mpie}{M_{\pi e}}
\newcommand{\mpimu}{M_{\pi\mu}}

\newcommand{\deltae}{\mbox{$\delta_K^e$}}
\newcommand{\deltam}{\mbox{$\delta_K^\mu$}}

\newcommand{\polep}{\mbox{$M_{v}$}}
\newcommand{\polez}{\mbox{$M_{s}$}}
\newcommand{\fplust}{\mbox{$\hat{f}_+(t)$}}
\newcommand{\fminust}{\mbox{$f_-(t)$}}
\newcommand{\fzerot}{\mbox{$\hat{f}_0(t)$}}
\newcommand{\tperp}{\mbox{$t_{\perp}$}}
\newcommand{\tperpb}{\mbox{$t^{\ell,\pi}_{\perp}$}}
\newcommand{\tperpp}{\mbox{$t^{\pi}_{\perp}$}}
\newcommand{\tperpl}{\mbox{$t^{\ell}_{\perp}$}}
\newcommand{\tperpe}{\mbox{$t^{e}_{\perp}$}}
\newcommand{\tperpm}{\mbox{$t^{\mu}_{\perp}$}}
\newcommand{\vus}{\mbox{$V_{us}$}}

\newcommand{\methodA}{fit-A}
\newcommand{\methodB}{fit-B}

\newcommand{\KLpienu}{\mbox{$K_{L}\to\pi^{\pm}e^{\mp}\nu$}}
\newcommand{\KLpilnu}{\mbox{$K_{L}\to\pi^{\pm}\ell^{\mp}\nu$}}
\newcommand{\KLpimunu}{\mbox{$K_{L}\to\pi^{\pm}{\mu}^{\mp}\nu$}}

\newcommand{\Gpienu}{\mbox{$\Gamma_{Ke3}$}}
\newcommand{\Gpimunu}{\mbox{$\Gamma_{K\mu3}$}}

 \newcommand{\beq}{\begin{equation}}
 \newcommand{\eeq}{\end{equation}}
 \newcommand{\bqa}{\begin{eqnarray}}
 \newcommand{\eqa}{\end{eqnarray}}

\def\epe{\epsilon'\!/\epsilon}

\newcommand{\ktev}{\mbox{KTeV}}

\def\kpi0{K_{L}\to 3\pi^0}
\def\ke3{K_{L}\to\pi^{\pm}e^{\mp}\nu}
\def\k2pi{K_{L} \to \pi^+\pi^-}

\newcommand{\Kethree}{\mbox{$K_{e3}$}}
\newcommand{\Kmuthree}{\mbox{$K_{\mu 3}$}}

\newcommand{\KLpm}{\mbox{$K_{L}\rightarrow\pi^{+}\pi^{-}$}}

\newcommand{\KLpmz}{\mbox{$K_{L}\rightarrow \pi^{+}\pi^{-}\pi^{0}$}}

\newcommand{\ring}{{\tt RING}}

\def\RPMNvalue{0.6640}
\def\RPMNerrstat{0.0014}
\def\RPMNerrsyst{0.0022}

\def\lplAv{28.32}
\def\lplaBv{21.67}
\def\lplbBv{2.87}
\def\lplCv{881.03}

\def\lzEv{16.57}
\def\lplEv{27.45}
\def\lzFv{12.81}
\def\lplaFv{17.03}
\def\lplbFv{4.43}
\def\lplGv{889.19}
\def\lzGv{1167.14}

\def\lplAe{0.37}
\def\lplaBe{1.37}
\def\lplbBe{0.57}
\def\lplCe{5.12}

\def\lzEe{0.78}
\def\lplEe{0.88}
\def\lzFe{1.36}
\def\lplaFe{3.19}
\def\lplbFe{1.31}
\def\lplGe{12.81}
\def\lzGe{28.30}

\def\lplAm{0.22}
\def\lplaBm{0.80}
\def\lplbBm{0.33}
\def\lplCm{2.96}

\def\lzEm{0.31}
\def\lplEm{0.35}
\def\lzFm{0.55}
\def\lplaFm{1.28}
\def\lplbFm{0.52}
\def\lplGm{5.14}
\def\lzGm{11.34}

\def\lplAt{0.57}
\def\lplaBt{1.99}
\def\lplbBt{0.78}
\def\lplCt{7.11}

\def\lzEt{1.25}
\def\lplEt{1.08}
\def\lzFt{1.83}
\def\lplaFt{3.65}
\def\lplbFt{1.49}
\def\lplGt{16.20}
\def\lzGt{42.00}

\def\lplAs{0.43}
\def\lplaBs{1.43}
\def\lplbBs{0.53}
\def\lplCs{4.94}

\def\lzEs{0.98}
\def\lplEs{0.63}
\def\lzFs{1.22}
\def\lplaFs{1.77}
\def\lplbFs{0.72}
\def\lplGs{9.92}
\def\lzGs{31.04}

\def\corB{-0.97}
\def\corE{-0.38}
\def\coraF{-0.96}
\def\corbF{0.65}
\def\corcF{-0.75}
\def\corG{-0.46}

\def\chA{81.0}
\def\chB{62.2}
\def\chC{66.3}

\def\chE{240.4}
\def\chF{230.7}
\def\chG{234.7}

\def\ndfA{65}
\def\ndfB{64}
\def\ndfC{65}

\def\ndfE{236}
\def\ndfF{235}
\def\ndfG{236}

\def\lplav{28.13}
\def\lzav{16.35}
\def\lplabv{20.64}
\def\lplbbv{3.20}
\def\lzbv{13.72}
\def\lplcv{882.32}
\def\lzcv{1173.80}

\def\lplae{0.51}
\def\lzae{1.21}
\def\lplabe{1.75}
\def\lplbbe{0.69}
\def\lzbe{1.31}
\def\lplce{6.54}
\def\lzce{39.47}

\def\cora{-0.36}
\def\corab{-0.97}
\def\corbb{0.34}
\def\corcb{-0.44}
\def\corc{-0.40}

\def\cha{0.5}
\def\chb{1.5}
\def\chc{0.2}

\def\ikeav{0.15507}
\def\ikmav{0.10294}
\def\ikebv{0.15350}
\def\ikmbv{0.10165}
\def\ikecv{0.15445}
\def\ikmcv{0.10235}

\def\ikeae{0.00027}
\def\ikmae{0.00026}
\def\ikebe{0.00044}
\def\ikmbe{0.00039}
\def\ikece{0.00023}
\def\ikmce{0.00022}

\def\ikrav{0.6639}
\def\ikrbv{0.6622}
\def\ikrcv{0.6627}

\def\ikrae{0.0017}
\def\ikrbe{0.0017}
\def\ikrce{0.0015}

\def\LeptUni{0.9969}
\def\LeptUniE{0.0048}
\def\DeltaRat{1.0058}
\def\DeltaRatE{0.0010}

\def\ikrbetot{0.0018}
\def\ikebetot{0.00105}
\def\ikmbetot{0.00080}
\def\ikebepar{0.00095}
\def\ikmbepar{0.00070}

\begin{document}

\title{
  Measurements of Semileptonic $K_L$ Decay Form Factors
 }

\newcommand{\UAz}{University of Arizona, Tucson, Arizona 85721}
\newcommand{\UCLA}{University of California at Los Angeles, Los Angeles,
                    California 90095} 
\newcommand{\UCSD}{University of California at San Diego, La Jolla,
                   California 92093} 
\newcommand{\EFI}{The Enrico Fermi Institute, The University of Chicago, 
                  Chicago, Illinois 60637}
\newcommand{\UB}{University of Colorado, Boulder, Colorado 80309}
\newcommand{\ELM}{Elmhurst College, Elmhurst, Illinois 60126}
\newcommand{\FNAL}{Fermi National Accelerator Laboratory, 
                   Batavia, Illinois 60510}
\newcommand{\Osaka}{Osaka University, Toyonaka, Osaka 560-0043 Japan} 
\newcommand{\Rice}{Rice University, Houston, Texas 77005}
\newcommand{\UVa}{The Department of Physics and Institute of Nuclear and 
                  Particle Physics, University of Virginia, 
                  Charlottesville, Virginia 22901}
\newcommand{\UW}{University of Wisconsin, Madison, Wisconsin 53706}

\affiliation{\UAz}
\affiliation{\UCLA}
\affiliation{\UCSD}
\affiliation{\EFI}
\affiliation{\UB}
\affiliation{\ELM}
\affiliation{\FNAL}
\affiliation{\Osaka}
\affiliation{\Rice}
\affiliation{\UVa}
\affiliation{\UW}

\author{T.~Alexopoulos}   \affiliation{\UW}
\author{M.~Arenton}       \affiliation{\UVa}
\author{R.F.~Barbosa}     \altaffiliation[Permanent address: ]
   {University of S\~{a}o Paulo, S\~{a}o Paulo, Brazil}\affiliation{\FNAL}
\author{A.R.~Barker}      \altaffiliation[Deceased.]{ } \affiliation{\UB}
\author{L.~Bellantoni}    \affiliation{\FNAL}
\author{A.~Bellavance}    \affiliation{\Rice}
\author{E.~Blucher}       \affiliation{\EFI}
\author{G.J.~Bock}        \affiliation{\FNAL}
\author{E.~Cheu}          \affiliation{\UAz}
\author{S.~Childress}     \affiliation{\FNAL}
\author{R.~Coleman}       \affiliation{\FNAL}
\author{M.D.~Corcoran}    \affiliation{\Rice}
\author{B.~Cox}           \affiliation{\UVa}
\author{A.R.~Erwin}       \affiliation{\UW}
\author{R.~Ford}          \affiliation{\FNAL}
\author{A.~Glazov}        \affiliation{\EFI}
\author{A.~Golossanov}    \affiliation{\UVa}
\author{J.~Graham}        \affiliation{\EFI}
\author{J.~Hamm}          \affiliation{\UAz}
\author{K.~Hanagaki}      \affiliation{\Osaka}
\author{Y.B.~Hsiung}      \affiliation{\FNAL}
\author{H.~Huang}         \affiliation{\UB}
\author{V.~Jejer}         \affiliation{\UVa}
\author{D.A.~Jensen}      \affiliation{\FNAL}
\author{R.~Kessler}       \affiliation{\EFI}
\author{H.G.E.~Kobrak}    \affiliation{\UCSD}
\author{K.~Kotera}        \affiliation{\Osaka}
\author{J.~LaDue}         \affiliation{\UB}
\author{A.~Ledovskoy}     \affiliation{\UVa}
\author{P.L.~McBride}     \affiliation{\FNAL}
\author{E.~Monnier}
   \altaffiliation[Permanent address: ]{C.P.P. Marseille/C.N.R.S., France}
   \affiliation{\EFI}
\author{H.~Nguyen}       \affiliation{\FNAL}
\author{R.~Niclasen}     \affiliation{\UB} 
\author{V.~Prasad}       \affiliation{\EFI}
\author{X.R.~Qi}         \affiliation{\FNAL}
\author{E.J.~Ramberg}    \affiliation{\FNAL}
\author{R.E.~Ray}        \affiliation{\FNAL}
\author{M.~Ronquest}	 \affiliation{\UVa}
\author{E. Santos}       \altaffiliation[Permanent address: ]
             {University of S\~{a}o Paulo, S\~{a}o Paulo, Brazil}
                          \affiliation{\FNAL}
\author{P.~Shanahan}     \affiliation{\FNAL}
\author{J.~Shields}      \affiliation{\UVa}
\author{W.~Slater}       \affiliation{\UCLA}
\author{D.~Smith}	 \affiliation{\UVa}
\author{N.~Solomey}      \affiliation{\EFI}
\author{E.C.~Swallow}    \affiliation{\EFI}\affiliation{\ELM}
\author{R.J.~Tesarek}    \affiliation{\FNAL}
\author{P.A.~Toale}      \affiliation{\UB}
\author{R.~Tschirhart}   \affiliation{\FNAL}
\author{Y.W.~Wah}        \affiliation{\EFI}
\author{J.~Wang}         \affiliation{\UAz}
\author{H.B.~White}      \affiliation{\FNAL}
\author{J.~Whitmore}     \affiliation{\FNAL}
\author{M.~Wilking}      \affiliation{\UB}
\author{B.~Winstein}     \affiliation{\EFI}
\author{R.~Winston}      \affiliation{\EFI}
\author{E.T.~Worcester}  \affiliation{\EFI}
\author{T.~Yamanaka}     \affiliation{\Osaka}
\author{E.~D.~Zimmerman} \affiliation{\UB}

\collaboration{The KTeV Collaboration}

\begin{abstract}
  We present new measurements of $K_L$ semileptonic form factors
  using data collected in 1997 by the 
  {\ktev}  (E832)  experiment at Fermilab. The measurements are based
  on $1.9$ million \KLpienu\ 
  and $1.5$ million \KLpimunu\ decays.
  For $f_+(t)$, we measure both
  a linear and quadratic term:
  $\lplusp = (\lplabv \pm \lplabe)\times10^{-3}$   
  and $\lpluspp = (\lplbbv \pm \lplbbe)\times10^{-3}$.
  For $f_0(t)$, we find $\lzero =   (\lzbv \pm \lzbe)\times10^{-3}$.
  These form factors are  consistent with $K^\pm$
  form factors, suggesting that
  isospin symmetry breaking effects are small. We use our  measured
  values of the form factors to evaluate the decay phase space integrals,
  $I^e_K =\ikebv \pm \ikebetot$ and $I^\mu_K =\ikmbv \pm \ikmbetot$,
  where errors include uncertainties arising from the 
  form factor parametrizations.
\end{abstract}


\maketitle


%
%

\section{Introduction}
Measurements of   semileptonic form factors 
provide 
unique information about the dynamics of  strong interactions. 
The 
form factors for the
decays \KLpienu\ and \KLpimunu\
also are needed to determine  the decay phase 
space integrals, $I^{e}_K$ and $I^{\mu}_K$,  and thus 
are important for measuring the
Cabibbo-Kobayashi-Maskawa~\cite{cabibbo,km} matrix element 
$|\vus|$~\cite{leut-roos,ktev_prl}. 
In this paper, we report a new measurement of the semileptonic 
form factors based on data collected by the KTeV experiment at Fermilab.

The dependence of the semileptonic form factors 
on $t$, the four-momentum transfer to the leptons squared,
was
under serious experimental and theoretical 
study from the late 1960s to the early 1980s; 
the result of this effort was the precise measurement of the form 
factor in the decay \KLpienu~\cite{pdg02}. Nearly twenty years later,
this  form factor  measurement  
was confirmed by 
CPLEAR~\cite{cplr00}
with precision comparable to the older experiments.
The measurement
 presented in this paper provides another check of the older
results with better precision.

The form factors in the decay \KLpimunu,
on the other hand,  have been a long standing
 subject of  experimental controversy~\cite{pdg82}. 
The inconsistency of different
experimental results leads to a large uncertainty in  $I^{\mu}_K$,
reducing the usefulness of the \KLpimunu\ partial width 
in the extraction of $|\vus|$.   
The new KTeV measurement  reduces the uncertainty
in $I^{\mu}_K$ to the same level as in $I^e_K$.
The improved
precision of $I^{\mu}_K$  
allows a sensitive test of the consistency between the phase
space integrals and the ratio of semileptonic partial widths
\Gpimunu/\Gpienu.

This paper is organized as follows. Section~\ref{sec:phen} introduces the 
phenomenology of semileptonic form factors.
Section~\ref{sec:teq}
describes the  experimental technique used
to determine the form factor values. The KTeV detector, Monte Carlo
simulation (MC), and data selection are described in
Section~\ref{sec:detector}. Section~\ref{sec:fit} explains
the fitting procedure to extract the form factors and
Section~\ref{sec:syst} describes the systematic uncertainties
in the analysis.
Finally, in Section~\ref{sec:results},
we present results for the form factors and for the 
phase space
integrals.

\section{Form Factor Phenomenology}
\begin{figure}
\centering
\psfig{figure=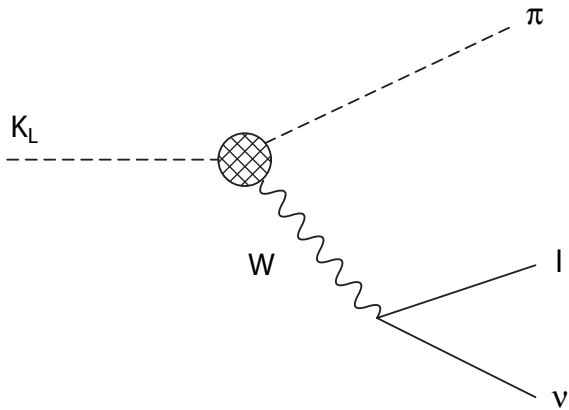,width=7.5cm}
\caption{Feynman diagram for the decay \KLpilnu. The hashed
circle indicates the hadronic vertex that depends
on decay form factors. }
\label{fig:diag}
\end{figure}
\label{sec:phen}
In the Standard Model, the Born-level matrix element for the \KLpilnu\ decay
modes (Fig.~\ref{fig:diag})
is~\footnote{Our notation in this paper is as follows:
four-vectors are denoted by $P_K$, $P_\pi$, etc. Spinors are denoted by
$u_\ell$, $u_\nu$, etc. Three-vectors are given in bold
letters. Variables in the kaon rest frame are denoted by $P^*_{\nu},
P^*_{\pi}$, etc. Components of three-vectors, parallel and perpendicular
to the kaon direction are denoted by $p_{\perp}$ and $p_{\parallel}$, respectively.}
\begin{equation}
 \begin{array}{r}
 {\cal M} = \frac{\textstyle G_F}{\textstyle \sqrt{2}} \vus 
      [
      f_+(t) (P_K + P_\pi)^\mu \overline{u_\ell} \gamma_\mu (1 + \gamma_5) u_\nu \\
     +\fminust (P_K - P_\pi)^\mu \overline{u_\ell} \gamma_\mu (1 + \gamma_5) u_\nu ],
 \end{array}
\label{eq:me}
\end{equation}
where $G_F$ is the Fermi constant and
$f_+$ and $f_-$ are the  vector form factors. 
Time-reversal invariance  guaranties that the form factors are real,
while local creation of the lepton pair requires that 
they be functions only of the square of the
 four momentum transfer to the leptons:
\begin{equation}
 Q^2 = (P_\ell + P_\nu)^2 = (P_K - P_\pi)^2 = t~.\label{eq:t}
\end{equation}
The above equality holds for  Born-level kinematics.
For  radiative $\KLpilnu\gamma$  
events, we assume that the form factors depend only on the virtuality
of the exchange boson at the hadronic vertex~\footnote{For radiative
$\KLpilnu\gamma$ decays we consider diagrams with radiation off the pion, 
 lepton, and  hadronic vertex. For radiation off the lepton and  
vertex, 
   $t=(P_\ell + P_\nu + k)^2=(P_K-P_\pi)^2$; for radiation off 
the pion,  $t=(P_\ell + P_\nu)^2$.  }.

$f_+$ and $f_-$  are not the only possible pair of  form
factors that can describe  semileptonic kaon decays. More recent
experiments have chosen to use $f_0$  instead of $f_-$:
\begin{equation}
 f_0(t) = f_+(t) + \frac{\textstyle t}{M^2_K - M^2_\pi} \fminust,
\end{equation}
where $M_K$ and $M_\pi$ are the kaon and pion mass.
$f_+$ and $f_0$ correspond to the $1^-$
(vector) and $0^+$ (scalar) exchange amplitudes, respectively. 
For  the Born-level matrix element,
$f_0$ is multiplied by the lepton mass $M_\ell$;
for \KLpienu, it 
 is negligible compared to the contribution of 
$f_+$.

Note that for the $f_+(t)$ and $f_0(t)$  form factors,
$f_+(0) = f_0(0)$; otherwise \fminust\ is divergent 
for $t\to 0$. To simplify  the 
notation, we use  normalized form factors,
$\hat{f}_{+,0}(t)= f_{+,0}(t)/f_+(0)$.

Historically, the normalized form factors 
$\fplust$ and $\fzerot$  were 
usually analyzed assuming a linear $t$
dependence:
\begin{equation}
\begin{array}{l}
\fplust = 1 + \lplus \frac{\textstyle t}{\textstyle M^2_\pi} \\
\fzerot = 1 + \lzero \frac{\textstyle t}{\textstyle
M^2_\pi}~.  \\
\end{array}
\end{equation}
This linear parametrization
 gave an adequate description of the experimental data.
With a larger data sample and better control of  systematic
 uncertainties, 
we study the second order term in $f_+$~\footnote{Our $\KLpimunu$ 
sample is insufficient to study a second order term in $f_0$.}:
\begin{equation}
\label{eq:lampp}
\fplust = 1 + \lplusp \frac{\textstyle t}{\textstyle M^2_\pi} +
\frac{\textstyle 1}{\textstyle 2} \lpluspp 
 \frac{\textstyle t^2}{\textstyle M^4_\pi}~.
\end{equation}
The presence of higher order terms is
 motivated by the pole model, in which the
$t$ dependence of $\fplust$ and $\fzerot$ is described by exchange of the
lightest vector and scalar $K^*$ mesons, respectively:
\begin{equation}
\begin{array}{l}
\fplust = \frac{\textstyle M^2_{v} }{\textstyle M^2_{v} - t  } \\
\fzerot = \frac{\textstyle M^2_{s} }{\textstyle M^2_{s} - t  }~. 
\end{array} \label{eq:pole}
\end{equation}
Pole models expect $M_{v} \approx 892$~MeV,
the mass of the lightest vector strange-meson,
and $M_{s} > M_{v}$.
 The vector meson dominance picture
 is supported by the recent results on   $\tau \to K \pi \nu$ decays
 showing a large enhancement of the $K^*(892)$ channel~\cite{cleo_tau,tau_opal,tau_aleph}.

From the Born-level
matrix element (Eq.~\ref{eq:me}), the \KLpilnu\ decay rate is
\begin{equation}
\vusdef~.  \label{eq:vus}
\end{equation}
 $S_{EW}$~\cite{sew} is the short-distance  radiative correction
that is the same for both modes and
$(1+\delta^\ell_K)$~\cite{Ginsberg,Troy}
are mode-dependent, long-distance radiative corrections.
The dimensionless decay phase space integrals, $I^{\ell}_K$, depend
on the decay form factors~\cite{leut-roos}:
\begin{equation}
I^\ell_K = \int dt ~\frac{\textstyle 1}{\textstyle M_K^8}~ \lambda^{3/2} 
 ~F(t,\fplust,\fzerot)~,  
\label{eq:integral}
\end{equation}
where
\begin{equation}
\label{eq:ft}
\begin{array}{cl}
F(t,\fplust,&\fzerot) =  \left( 1+ \frac{\textstyle M_{\ell}^2}{\textstyle 2t} \right)
\left( 1 - \frac{\textstyle M_{\ell}^2} {\textstyle t} \right)^2 \times
 \\
& \left(
 \hat{f^2_+}(t) + \frac{\textstyle 3M_{\ell}^2
\left(M_K^2-M_\pi^2\right)^2}{\textstyle \left( 2t+M_{\ell}^2 \right)\lambda}
\  \hat{f^2_0}(t)
\right)
\end{array}
\end{equation}
and
\begin{equation}
\lambda = t^2 + M^4_K + M^4_\pi - 2tM^2_K - 2tM^2_\pi - 2M_K^2M_\pi^2~.
\end{equation}

\section{Form Factor Extraction Procedure}
\label{sec:teq}
For a  semileptonic decay detected by KTeV, 
the kaon direction of flight  is well
measured, but 
the kaon energy is not uniquely determined. 
Using the reconstructed missing transverse
momentum squared associated  with the neutrino  ($p^2_{\perp,\nu}$) 
and the 
 invariant mass of the pion-lepton system
($\mpilep$), one can uniquely determine the square of the neutrino
longitudinal momentum in the kaon rest frame:
\begin{equation}
  P^*_{\parallel,\nu}\-^2  = \frac{\textstyle \left(M^2_K - \mpilep
  \right)^2}{\textstyle 4 M^2_K} - p^2_{\perp,\nu}~.
\end{equation} 
The sign ambiguity for $P^*_{\parallel,\nu}$ leads to a twofold ambiguity for the
parent kaon energy, $E_K$, which in turn
gives rise to two different values of $t$. 

Most  previous fixed target experiments encountered the same 
twofold kaon energy ambiguity. 
Typical solutions to this problem included
selecting  events with small $|P^*_{\parallel,\nu}|$, 
picking the more probable solution,
or  using  both solutions with weights proportional to the
probability of each solution. In these approaches,
the probability of a given solution depended on
the form factors and the
kaon energy spectrum, which was 
often not  well understood; this coupling
increased the systematic uncertainty.

In this paper, we describe a technique to measure form factors
that avoids the twofold
$E_K$ ambiguity.
The key idea in this method is 
that instead of $t$, we use 
``transverse $t$,'' which is defined by substituting
the measured transverse momenta for the  
full $3$-momenta in Eq.~\ref{eq:t}. A similar approach was 
used in~\cite{Brandenburg73}.

We introduce two definitions of transverse $t$
following the two definitions of $t$. We define ``lepton transverse
$t$'', \tperpl, by modifying 
$(P_{\ell} + P_{\nu})^2$:
\begin{equation}
\begin{array}{lcl}
\tperpl  &= & M^2_{\ell} + 2  |p_{\perp,\nu}| \sqrt{p^2_{\perp,\ell} + M^2_{\ell}} - 
2{\bf p_{\perp,\nu} p_{\perp,\ell}}   \label{eq:tperp}\\
           & = & M^2_{\ell} + 2  |p_{\perp,\nu}| \sqrt{p^2_{\perp,\ell} + M^2_{\ell}}
           + p_{\perp,\ell}^2 + p_{\perp,\nu}^2 - p_{\perp,\pi}^2~. \\
\end{array}
\end{equation}
In a similar way, we define ``pion transverse $t$,'' \tperpp, by
modifying 
$(P_K - P_{\pi})^2$:
\begin{equation}
\tperpp  = M^2_K + M_\pi^2 - 2 M_K \sqrt{p^2_{\perp,\pi} + M_\pi^2}~. 
\end{equation}
Note that from Eq.~\ref{eq:t}, the following inequality holds:
\begin{equation}
       \tperpl \le t \le \tperpp~. \label{eq:ineq}
\end{equation}
To simplify the notation,  we use the symbol
$\tperp$ if the discussion refers to either $\tperpp$ or
$\tperpl$; the symbol $\tperpb$ is used to refer to
both $\tperpp$ and $\tperpl$.

Since the $K_L$ is a scalar particle, the decay kinematics does not depend
on the $K_L$ direction. Therefore, the \tperp\ distribution can be
unambiguously related to the $t$ distribution and vice versa. 
The value of \tperp\ is invariant under  kaon boosts, and thus the \tperp\ distribution
does not
depend on the kaon energy spectrum. The loss of longitudinal information
does not dramatically increase the statistical uncertainty: 
a MC study shows that 
the \tperp-based measurement of the form factors leads to a
$15\%$ 
increase in  uncertainty 
with respect to an ideal, $t$-based measurement.

Each  definition of $t_\perp$
can be used to measure the form factors. 
We will refer to the extraction of  form factors using the \tperpl\ 
or \tperpp\ variables as  the $\tperpl$- or $\tperpp$-methods, respectively.
The two methods have a different
sensitivity to  systematic effects.
For example, the $\tperpp$-method is much less sensitive to 
radiative corrections, and therefore is used for the \KLpienu\ 
decay mode, while
the $\tperpl$-method leads to better statistical precision,
 and  is used for  the \KLpimunu\ decay mode.

The  form factor measurement starts with a determination of 
the \tperp\ distribution in data. We use a Monte 
Carlo simulation to determine
the detector acceptance as a 
function of \tperp\ and also to calculate  radiative
corrections. We extract the form factors by fitting the Monte Carlo 
simulation to data as  will be described in Section~\ref{sec:fit}.

\section{Apparatus, Simulation, and Data selection}
\label{sec:detector}
\begin{figure}
\centering
\psfig{figure=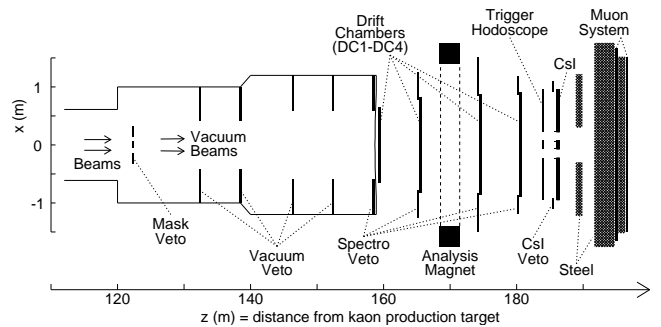,width=8.5cm}
\caption{Plan view of the KTeV (E832) detector.
The evacuated decay volume ends with a thin vacuum window at
$Z=159$~m.  The label ``CsI'' indicates the electromagnetic
calorimeter.}
\label{fig:detector}
\end{figure}

The KTeV experiment (Fig.~\ref{fig:detector})
and associated event reconstruction techniques have 
been described in detail elsewhere~\cite{Alavi-Harati:2002ye}. 
An 800 GeV/c proton beam striking a BeO target is used to produce two
almost parallel neutral beams. 
A large vacuum decay region surrounded by
photon veto detectors extends from 110 m to 159 m from the primary target. Following
a thin vacuum window at the end of the vacuum region is a drift chamber
spectrometer, equipped with an analysis magnet 
which imparts a 0.41 GeV/c 
 kick in the horizontal plane.
Farther downstream, there is a 
trigger hodoscope, a pure CsI electromagnetic calorimeter,
and a muon system 
consisting of two scintillator hodoscopes behind 4 m 
and 5 m of steel.

A detailed Monte Carlo simulation is used to correct for the acceptance
of the \KLpilnu\ decay modes. 
The MC includes three main steps: event generation,
propagation of particles through the detector,
and simulation of the detector performance.  
A thorough discussion of 
this simulation program is given in~\cite{Alavi-Harati:2002ye},
and more specifically for   \KLpilnu\ decays  in~\cite{ktev_kbr}.
To include leading-order QED radiative corrections, we use the KLOR
program~\cite{Troy}.

For this analysis, 
we use a low intensity data sample collected in 1997, 
which is the same data sample used to measure 
 $\Gamma(\KLpimunu)/ \Gamma(\KLpienu)$~\cite{ktev_kbr}. 
Consequently, most of the data selection
requirements are similar to those used in the
KTeV partial width ratio analysis.

The reconstruction begins with the identification of two oppositely charged tracks
coming from a single vertex, reconstructed between $123$ and $158$~m
from the primary target. 
To pass the event selection,  one of the two tracks must
 be within $7$~cm  of a cluster in the CsI calorimeter; the second
track is not required to have a cluster match. The event's
missing transverse momentum squared  associated with the 
neutrino
is calculated with respect to the line connecting the primary
target and the decay vertex. Both kaon energy solutions 
are required to fall between $40$ and $120$~GeV, where the
kaon momentum spectrum is well measured.

The decay \KLpienu\ is identified using $E/p$, 
the energy reconstructed in the CsI calorimeter 
divided by the momentum measured in the spectrometer.
One track is required to have $E/p$ greater than
 0.92
and the other track  is required  
to have $E/p$ less than 0.85. The first condition 
is satisfied for $99.8\%$ of electrons and failed by $99.5\%$ of pions.
The second condition is satisfied for  $99.1\%$ of  pions and
failed by  $99.93\%$ of electrons. The resulting unambiguous 
identification of the pion and electron in 
the decay  \KLpienu\  (swap probability $<0.01\%$) allows
for the correct reconstruction of $\tperp$. 

The decay \KLpimunu\ is identified by requiring one of the tracks
to have an energy deposit in the CsI less than 
$2$~GeV~\footnote{A minimum ionising particle deposits about 0.4~GeV in the 50~cm
long CsI crystals.}, and the other track
to have $E/p$ less than 0.85. This selection strongly suppresses 
\KLpienu\ decays;
however, since the CsI calorimeter is only about 1/3 of a hadronic
interaction length, the calorimeter selection is not very effective 
in reducing background from
\KLpmz\ and \KLpm.

In both semileptonic decay modes,
\KLpmz\ background is rejected using  the variable \ppzkin~\cite{ktev_kbr}.
For a genuine \KLpmz\ event, \ppzkin\ is proportional to the square
of the  $\pi^0$ longitudinal momentum in the reference frame in
which  the sum of $\pi^+$-$\pi^-$ momenta is orthogonal to the kaon
direction.
We require $\ppzkin < -0.006$,  rejecting $99.9\%$ of \KLpmz\ decays
while retaining $99.5\%$ of \KLpienu\ and 
$97.9\%$ of \KLpimunu\ decays. \KLpm\ background is suppressed
by rejecting events in which the two-track invariant mass, assuming both 
tracks are pions, is between $0.488$ and $0.505$~GeV/c$^2$.

The main difference in the selection of the decay \KLpimunu\ 
compared to the partial width ratio analysis is driven by 
the need to distinguish between the pion and  muon tracks
in order to reconstruct $\tperp$. 
Since the calorimeter selection does not uniquely identify the pion
and muon, we identify the muon
by requiring the extrapolation of a track
to be near  fired horizontal and vertical counters in the muon  system.
The quality of this ``match'' is measured by a  $\chi^2_{\mu}$
variable that accounts for  momentum-dependent
multiple scattering in the steel in front of the muon system.
We reject 
events with two matched 
tracks to suppress muon misidentification from $\pi \to \mu\nu$
decays and from pion penetrations through the steel. The misidentification
is suppressed further by requiring 0.5~m separation between the two tracks
at the muon system (compared to the $0.15$~m segmentation 
of the muon system counters). The efficiency of the
matching requirement is increased by demanding that the muon track projection
 be at least $0.25$~m inside the muon system outer aperture.
To reduce the fraction of muons ranging out in the steel,
we also require that  the muon track momentum  be greater than 
$10$~GeV.
These selection criteria result in the unambiguous identification of
the pion and  muon in the decay \KLpimunu, with a swap probability
of less than $0.1\%$.

There are two categories of background in this analysis: misidentified
kaon decays and
scattering 
background
caused by a kaon scattering off a beamline element
(absorber or collimator).
The misidentification background for the \KLpienu\ decay mode is below 
 $10^{-5}$  and is ignored.
The background for the  \KLpimunu\ decay mode is also
small ($ 10^{-4}$), but  is not uniform in \tperpb\ 
and must be subtracted.
Kaon scattering off  beamline 
elements, resulting in an incorrect measurement of $p^2_\perp$, 
affects $(0.11\pm 0.01) \%$ of all events.
Systematic uncertainties arising from the modeling of the 
background are discussed in Section~\ref{sec:bg}.

\begin{figure}
\centering
\psfig{figure=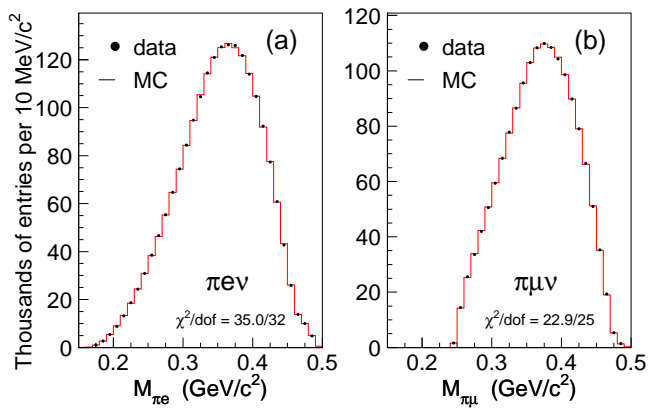,width=\linewidth}
\caption{The invariant mass distribution (a) $\mpie$ for
\KLpienu\ events  and (b)  $\mpimu$ for \KLpimunu\ events. 
Data are shown as dots; MC is shown as a histogram.
All analysis requirements have been applied.
\label{fig:mass}}
\end{figure}

\begin{figure}
\centering
\psfig{figure=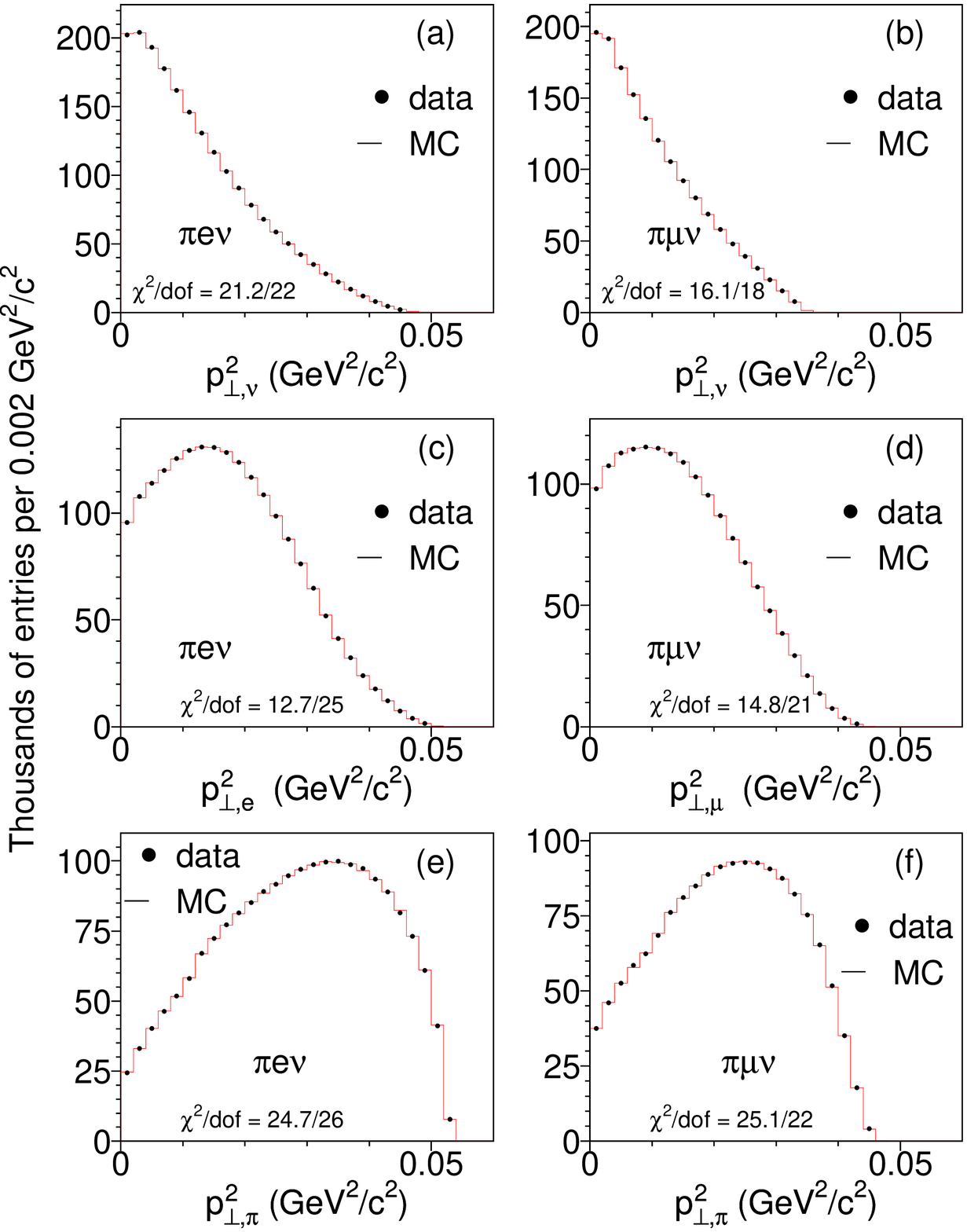,width=\linewidth}
\caption{Distribution of 
(a) $p^2_{\perp,\nu}$  for \KLpienu\ events, 
(b) $p^2_{\perp,\nu}$  for \KLpimunu\ events, 
(c) $p^2_{\perp,e}$  for \KLpienu\ events, 
(d) $p^2_{\perp,\mu}$ for \KLpimunu\ events, 
(e) $p^2_{\perp,\pi}$ for \KLpienu\ events, and 
(f) $p^2_{\perp,\pi}$ for \KLpimunu\ events (d). 
Data are shown as dots, signal MC is shown as histogram.
All analysis requirements have been applied.
\label{fig:pt}}
\end{figure}

The main kinematic variables for the extraction of the
semileptonic form factors are the transverse momenta-squared
of all three particles, $p^2_{\perp,\nu}$,
 $p^2_{\perp,\ell}$
and $p^2_{\perp,\pi}$, as well as the pion-lepton invariant mass,  
$\mpilep$.
Distributions of  $\mpilep$
for the \KLpienu\ and \KLpimunu\ decay modes are 
shown in Figure~\ref{fig:mass}.
Distributions 
of $p^2_{\perp,\nu}$, $p^2_{\perp,\ell}$ and  $p^2_{\perp,\pi}$ are shown in 
Figure~\ref{fig:pt}. 

Note that the $\mpilep$ and $p_\perp^2$ distributions 
(Figs.~\ref{fig:mass},\ref{fig:pt}) have different
sensitivity to radiative  effects and form factor values. 
For example, using the \KLpienu\ MC without radiative effects,
the form factor  obtained
with the $\tperpp$-method gives good data-MC agreement in $p_{\perp,\pi}^2$,
but poor agreement in the $\mpie$ and $p_{\perp,e}^2$
distributions. Conversely, the form factor obtained  
with the \tperpl-method gives good
data-MC agreement in
$\mpie$ and $p_{\perp,e}^2$, but  poor
agreement in $p_{\perp,\pi}^2$. 
The excellent data-MC agreement shown in  all four distributions 
($\mpie$, $p_{\perp,\nu}^2$, $p_{\perp,e}^2$ and $p_{\perp,\pi}^2$ in Figs.~\ref{fig:mass},\ref{fig:pt}) is possible only if radiative
effects are included in the MC.

The measurement of the semileptonic form factors benefits
greatly from the excellent resolution of the KTeV spectrometer
($<1\%$), and also from the extensive calibration performed
for the measurement of $\epe$~\cite{Alavi-Harati:2002ye}. 
The resulting  $\tperp$ resolution
is better than $1.5\%$ on average.

\section{Fitting}
\label{sec:fit}
\begin{figure}
\centering
\psfig{figure=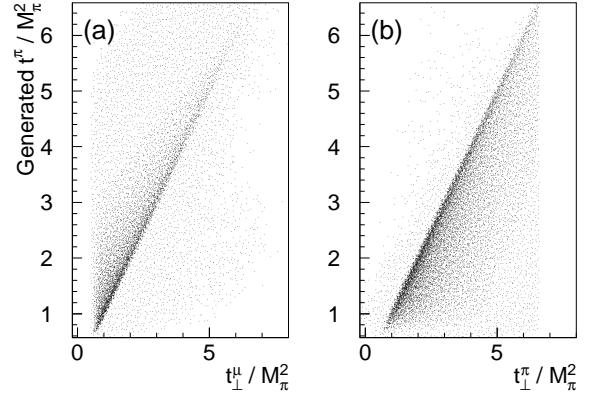,width=\linewidth}
\caption{\label{fig:corel}The distribution of
generated $t^\pi$ versus
reconstructed \tperp, which is used
to construct the matrix ${\cal A}_{rg}$, for
(a) \tperpm-method and 
(b) \tperpp-method. 
The plot is based on the \KLpimunu\ decay mode.}
\end{figure}

We employ two different strategies to extract the form factors
from the measured  \tperp\ distribution. In the first method
(``\methodA''),
we use the  MC to obtain the matrix
${\cal A}_{rg}$, which relates the {generated}
$t^\pi=(P_K - P_{\pi})^2$, $g$-index, 
to the {reconstructed} \tperpl or \tperpp, $r$-index.  
Note that detector resolution  and radiative effects
  are included in this matrix.
Figure~\ref{fig:corel} shows the distribution
of $t^\pi$ and \tperpb\ that is
used to calculate the matrix ${\cal A}_{rg}$~\footnote{Entries 
below the diagonal in Fig.~\ref{fig:corel} (a)  and
above the diagonal in Fig.~\ref{fig:corel} (b) fail the condition
stated in Eq.~\ref{eq:ineq}. These
unphysical entries are due to misreconstruction 
caused by  resolution effects and scattering.  }.

Using the matrix ${\cal A}_{rg}$, we construct the  prediction
for the binned distribution 
of the reconstructed \tperp:
\begin{equation}
   {\cal N}^{predA}_{\perp,r} = N_0 \sum_g {\cal A}_{rg} ~
  \frac{\textstyle F(t_g, \hat{f}^{pred}_+(t_g),\hat{f}^{pred}_0(t_g)) } 
  {\textstyle F(t_g, \hat{f}^{MC}_+(t_g),\hat{f}^{MC}_0(t_g))}
  \label{eq:fita}.
\end{equation}
Here, $F(t,\fplust,\fzerot)$ is the
Born-level $t$ distribution function (Eq.~\ref{eq:ft})
which depends on the form factors~\footnote{
Radiative corrections for the $t^{\pi}$ distribution
are small and depend weakly on the form factors. Radiative corrections
 therefore
can be neglected in the ratio of 
$F(t_g, \hat{f}^{pred}_+(t_g),\hat{f}^{pred}_0(t_g))$ to
$F(t_g, \hat{f}^{MC}_+(t_g),\hat{f}^{MC}_0(t_g))$ if the MC  
uses form factors which are reasonably close to the true values.}; 
$\hat{f}_{+,0}^{MC}$ are the form factors
used in MC and $\hat{f}^{pred}_{+,0}$ are the form factors floated in the
fit; $t_g$
are the bin centers and $N_0$ is the overall normalization, which is
also floated in the fit.
For $\hat{f}^{pred}_{+,0}=\hat{f}_{+,0}^{MC}$, the prediction 
${\cal N}^{predA}_{\perp,r}$ is simply the reconstructed  \tperp\ distribution
from MC.
For different $\hat{f}^{pred}_{+,0}$, the prediction function
(Eq.~\ref{eq:fita})
is  equivalent to ``reweighting'' the MC to the new
values of the form factors.
The actual fit is the $\chi^2$ minimization between the binned \tperp\
distribution in data and the prediction ${\cal N}^{predA}_{\perp,r}$.

In the second method (``\methodB''),
we follow a  common approach in which 
we fit the  data \tperp\
distribution, corrected for  
acceptance and radiative effects,
to the Born-level prediction.
First, we
define the acceptance $A_b$ in a given bin of  $t_{\perp,b}^{\ell,\pi}$ as the 
ratio of the number of
reconstructed events to the number of  generated events in the bin. Next, 
we  define the radiative correction $(1+\delta_b)$ for each bin
as the ratio of the number of 
generated events for the nominal MC (including radiative
effects from KLOR) to the number of  generated
events for a separate Born-level MC; these MC samples are normalized using the 
global radiative correction $(1+\delta^\ell_K)$. 
After correcting for acceptance and radiative effects,
the number of corrected data events
for a given $\tperp$ bin is given  by 
\begin{equation}
  N^{Born}_{\perp,b} = \frac{\textstyle N^{rec}_{\perp,b} }
 {\textstyle A_b ( 1 + \delta_b )}~,
\end{equation}
where $N^{rec}_{\perp,b}$ is the number of reconstructed events.
Both acceptance and radiative corrections
depend on the form factors and are therefore determined in an iterative
procedure starting with the PDG values.
Since  this form factor  dependence is relatively
weak, only one iteration is required for  convergence.

To determine the prediction for the
Born-level \tperp\ distribution, we start with a quadratic parametrization
of the form factors (Eq.~\ref{eq:lampp}). 
We square the matrix element
(Eq.~\ref{eq:me}) and collect terms proportional to 
$\lplusp,\lplusp^2,\lpluspp,\lpluspp^2$, 
$\lzerop,\lzerop^2,\lzeropp,\lzeropp^2$, and cross terms.
As a result,  $|{\cal M}|^2$ may be expressed as a second order function
of $\lplusp,\lzerop,\lpluspp,\lzeropp$:
 \begin{equation}
  |{\cal M}|^2 = {\cal I}
 + \lplusp {\cal I}^{+'}
 + \lzerop {\cal I}^{0'}
 + ... \label{eq:fff}~,
 \end{equation}
where ${\cal I}$ are  functions of the Dalitz plot variables.
We generate \KLpienu\ and \KLpimunu\ decays according to
 three-body phase space kinematics and numerically
 integrate over  ${\cal I}$ for
events falling into the same  $\tperp$ bin. 
In this way, the prediction 
for the binned \tperp\ distribution can be expressed as
\begin{equation}
  {\cal N}^{predB}_{\perp,b}(\lplusp,\lzerop,...) = {\cal I}_{\perp,b} 
 + \lplusp {\cal I}^{+'}_{\perp,b} 
 + \lzerop {\cal I}^{0'}_{\perp,b} 
 + ...~, \label{eq:fitb}
\end{equation}
where ${\cal I}_{\perp,b}$ are the results of the 
integration explained above.
For the pole model fit, we first Taylor expand 
the pole parametrization (Eq.~\ref{eq:pole}) up to second order in $t$ 
to obtain 
\begin{equation}
 \label{eq:taylor}
 \lambda'_{+} = \frac{M^2_{\pi}}{M^2_{v}}~~\mbox{and}~~
 \lambda''_{+} = 2 \frac{M^4_{\pi}}{M^4_{v}} 
\end{equation}
(and similar expressions for $\lzerop$ and $\lzeropp$ as a function of $M_s$),
and then use the prediction given by Eq.~\ref{eq:fitb}. 

\begin{figure}
\centering
\psfig{figure=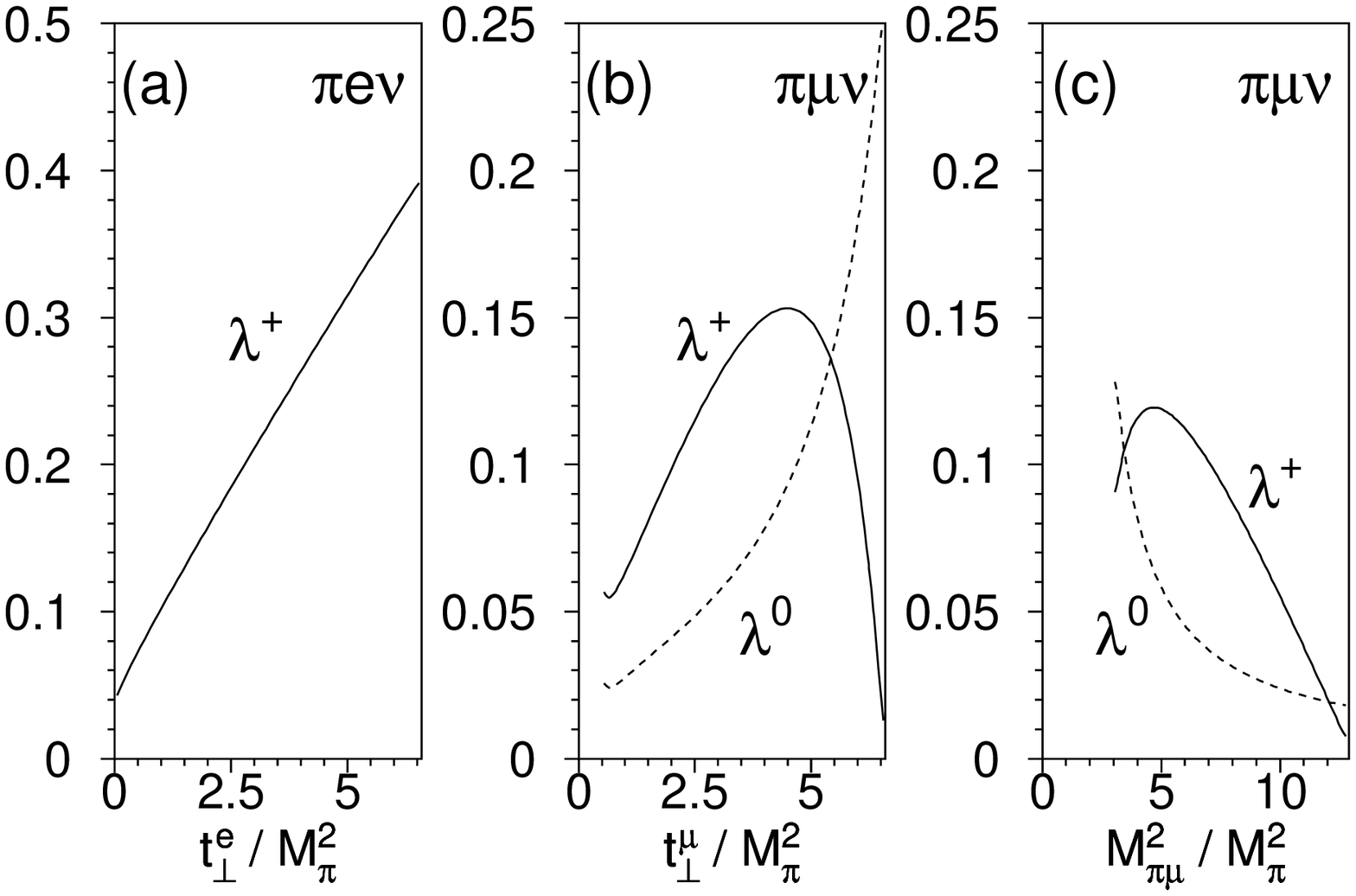,width=\linewidth}
\caption{
\label{fig:lamdepl}
Illustration of the influence of 
 the form factors $\lplus=0.03$ and $\lzero=0.02$ on the
$\tperpl$ and $\mpimu$ distributions. The figure shows
$0.03 \times {\cal I}^{+'}_{\perp,b}/{\cal I}_{\perp,b}$ (solid lines marked as
``$\lplus$'') and  $0.02 \times  {\cal I}^{0'}_{\perp,b}/ {\cal
I}_{\perp,b}$ (dashed lines marked as ``$\lzero$'') (a) as a function
of $\tperpe$ for decay \KLpienu, (b) as a function of
$\tperpm$ for decay \KLpimunu, and (c) as a function of $\mpimu$
for decay \KLpimunu.  
}
\end{figure}

The ratios  $\lplus  {\cal I}^{+}_{\perp,b}/{\cal I}_{\perp,b}$ 
and  $\lzero  {\cal I}^{0}_{\perp,b}/{\cal I}_{\perp,b}$
govern the influence of the \lplus\ and \lzero\ parameters on 
the $\tperpb$ distributions.
These ratios are illustrated in Figure~\ref{fig:lamdepl} for  typical
values of $\lplus$ and $\lzero$. In a similar manner, we determine
the influence of  \lplus\ and \lzero\  on the
 $\mpimu$ distribution
for the \KLpimunu\ decay mode (Figure~\ref{fig:lamdepl} (c)).
For the decay \KLpienu, $\lplus=0.03$ corresponds to a linear
increase of the decay rate as a function of $\tperpl$; for 
$\tperpl/M_{\pi^2}= 6$, this increase is about $40\%$. For 
\KLpimunu,  the effect of $\lzero$ is non-linear and is
more pronounced for large 
$\tperpl$ and small $m^2_{\pi\mu}$.

The advantage of  \methodA\ is that there is no need 
for an iterative determination of the acceptance
since the method is almost independent of the
form factors used in the MC. The matrix ${\cal A}_{rg}$
also utilizes the detector resolution information,
which leads to better statistical
precision compared to  \methodB\ (by about 10\%).
On the other hand, 
the  advantage of  \methodB\ is an easy generalization
to 2-dimensional fits, in particular to fit in $(\mpilep,\tperp)$ space.
Although \methodB\ has less precision than \methodA\ for
1-dimensional fits, the precision is improved in 2-dimensional
fits by using the different
shapes of the $\lzero$ and $\lplus$ distributions as a function of 
$m^2_{\pi\mu}$ (Fig.~\ref{fig:lamdepl} (c)).

Both fitting methods are checked using MC samples generated with different
values of form factors. The estimation  
of the statistical uncertainty is verified by generating 100 independent
MC samples and studying the distribution of the fitted form factors.
For the one dimensional fits in $\tperp$, 
the form factors for \KLpienu\ and \KLpimunu\ are extracted 
using both fitting techniques; these measurements are consistent within
the uncorrelated statistical uncertainty.

\section{Systematic Uncertainties}
\label{sec:syst}
The sources of systematic uncertainty for this analysis are the same
as for the partial width ratio measurements~\cite{ktev_kbr},
 but their impact on the result
is different. 
 The form factor measurement does not depend on uniform losses that do not
bias kinematic distributions. For example,  losses caused
by hadronic interactions inside the spectrometer have virtually no influence.
On the other hand,  effects that cause  misreconstruction of event
kinematics, such as calibration of the analysis magnet kick,
are more significant. 

Table~\ref{tab:syst} summarizes the systematic 
uncertainties for the form factor measurements.
For each decay mode and for each
type of  parametrization,
the uncertainties are evaluated for the results obtained using 
both the \tperpp\ and \tperpl\ methods. For the nominal KTeV result,
we select the method which leads to the smallest total error and use
the other method as a crosscheck.
The nominal form factor result for \KLpienu\  is based 
on a 1-dimensional fit
to the \tperpp\ distribution; the nominal form 
factor result for  \KLpimunu\  is based
on a 2-dimensional fit to the  ($\mpimu$,\tperpl)  distribution.

The systematic uncertainty in the
form factor resulting from  a given effect is evaluated by
turning this effect off in the MC simulation
and determining the corresponding 
change in the form factors. The systematic
uncertainty 
in the form factor is taken as this change 
times the relative uncertainty of the systematic effect. 
For example, turning off
muon scattering in the steel changes $\lzero$ by $1.5\times10^{-3}$;
since we model the effects of scattering with $20\%$ precision (Sec.~\ref{sec:muon}),
the corresponding uncertainty on $\lzero$ is $0.3\times 10^{-3}$. 

\begin{table*}
\centering
\caption{Systematic, statistical and total uncertainties
for the form factors.\label{tab:syst}}
\begin{tabular}{l|c|cc|c||cc|ccc|cc}
\hline
\hline
          & \multicolumn{4}{c||}{\Kethree}  & \multicolumn{7}{c}{\Kmuthree} \\
   Source & \lplus & \lplus' & \lplus'' & \polep & \lplus & \lzero & \lplus' & \lplus'' & \lzero & \polep & \polez \\
              &   \multicolumn{3}{c| }{$\times 10^{-3}$}  &    MeV      &         \multicolumn{5}{c|  }{$\times 10^{-3}$}                                     & MeV                &  MeV        \\
\hline
 Calibration \& Alignment& & & & & &  & & & & & \\
                         - Drift Chamber Alignment &   0.0 &   0.2 &   0.1 &    0. &   0.3 &   0.2 &   0.5 &   0.2 &   0.2 &    5. &    6. \\ 
                      - $P_\perp$ Kick Calibration &   0.2 &   1.0 &   0.3 &    2. &   0.2 &   0.3 &   0.0 &   0.1 &   0.4 &    3. &    9. \\ 
 Detector Simulation & & & & & &  & & & & & \\
                       - Multiple Scattering and Resolution &   0.1 &   0.5 &   0.1 &    2. &   0.0 &   0.1 &   0.3 &   0.1 &   0.1 &    0. &    2. \\ 
                          - Muon System Simulation &   0.0 &   0.0 &   0.0 &    0. &   0.1 &   0.3 &   0.3 &   0.1 &   0.3 &    2. &    9. \\ 
                                  - Cut Variations &   0.2 &   0.3 &   0.1 &    1. &   0.2 &   0.3 &   0.5 &   0.2 &   0.3 &    4. &   11. \\ 
             - Beam Shape and Kaon Energy Spectrum &   0.1 &   0.0 &   0.0 &    0. &   0.0 &   0.0 &   0.0 &   0.0 &   0.0 &    0. &    0. \\ 
 Background & & & & & & & & & & &  \\
                           - Scattering Background &   0.0 &   0.0 &   0.0 &    0. &   0.0 &   0.0 &   0.0 &   0.0 &   0.0 &    0. &    0. \\ 
                    - Misidentification Background &   0.0 &   0.0 &   0.0 &    0. &   0.1 &   0.1 &   0.4 &   0.1 &   0.0 &    2. &    3. \\ 
                                           Fitting &   0.1 &   0.2 &   0.1 &    1. &   0.2 &   0.2 &   0.4 &   0.3 &   0.2 &    3. &   10. \\ 
                             Radiative Corrections &   0.1 &   0.0 &   0.1 &    2. &   0.2 &   0.7 &   0.7 &   0.2 &   0.9 &    3. &   20. \\ 
  Monte Carlo statistics &\lplAm &\lplaBm & \lplbBm&\lplCm & \lplEm &\lzEm & \lplaFm&\lplbFm & \lzFm& \lplGm &  \lzGm \\
\hline
  Total systematic  & \lplAs  & \lplaBs  & \lplbBs & \lplCs &\lplEs & \lzEs  & \lplaFs &\lplbFs & \lzFs & \lplGs &  \lzGs  \\
\hline 
  Data statistics  & \lplAe  & \lplaBe  & \lplbBe & \lplCe &\lplEe & \lzEe  & \lplaFe &\lplbFe & \lzFe & \lplGe &  \lzGe  \\
\hline
  Total uncertainty  & \lplAt  & \lplaBt  & \lplbBt & \lplCt &\lplEt & \lzEt  & \lplaFt &\lplbFt & \lzFt & \lplGt &  \lzGt  \\
\hline
\hline
\end{tabular}
\end{table*}

The following sections discuss systematic uncertainties
in  the order given in Table~\ref{tab:syst}.

\subsection{Calibration \& Alignment}
\subsubsection{Drift chamber alignment}
The dimensions of the drift chambers 
are known to $20~\mu$m
from optical survey. The chamber locations along the
kaon beam ($z$-direction) were monitored during the run and 
are known to $100~\mu$m.
In the transverse directions, the chambers are aligned in-situ 
using dedicated muon runs with the analysis magnet turned
off. 
The transverse position of the primary target is known 
to $50~\mu$m precision by
projecting the total momentum of \KLpm\ events to  $Z=0$.

The alignment in the horizontal
 direction is most important because 
it affects the track momentum measurement. The residual uncertainty
in the alignment procedure is  $20~\mu$m. 
The systematic uncertainty reported in Table~\ref{tab:syst} is based
on the sum in quadrature of the form factor changes if 
DC1 is shifted  horizontally by $20~\mu$m, rotated in the $XY$ plane
by $20~\mu$rad, has non-orthogonality
between vertical and horizontal wires of $50~\mu$rad, or is shifted in $Z$
by $100~\mu$m.

\subsubsection{Magnet Kick Calibration}
The calibration of the analysis magnet transverse kick is performed
with \KLpm\ 
decays.
The $30$~keV uncertainty
in the PDG value of $M_K$  leads to a $0.01\%$ uncertainty
in the transverse kick.

\subsection{Detector Simulation}
\subsubsection{Multiple Scattering and Resolution}
As explained in~\cite{ktev_kbr},  the amount of detector material 
and the influence of electromagnetic interactions 
(multiple scattering, bremsstrahlung and $\delta$-ray production) 
on the detector acceptance
are described to $10\%$.
For the systematic uncertainty arising from modeling  of the drift chamber
position resolution, 
we assume a $5\%$ uncertainty for the Gaussian part of the 
chamber response and a $10\%$ uncertainty 
for the non-Gaussian tail induced by the discrete
statistics of the ionization process.  This estimate is based on a 
data-MC comparison of the distance between the track
and reconstructed hits. 
The chamber inefficiency caused by  early accidental
activity as well as hadronic interactions are negligible for this analysis. 

\subsubsection{Muon System Simulation}
\label{sec:muon}
Several effects must be  simulated to reproduce  the
efficiency of the muon system.
These effects are  muon
scattering in the steel, modeling of the cracks between the muon 
system counters, and
pion penetrations through the steel.

\begin{figure}
\centering
\psfig{figure=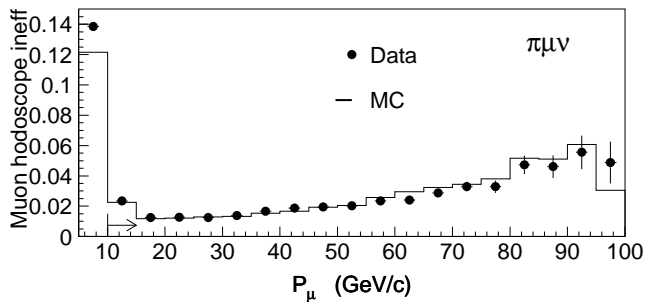,width=\linewidth}
\caption{\label{fig:peff}
Inefficiency of the muon hodoscope matching requirement 
as a function of the muon momentum
measured with \KLpimunu\ decays. Arrow indicates
the momentum requirement. The inefficiency increases with
momentum because of the $\chi^2_{\mu}$ definition. 
}
\end{figure}

Multiple scattering in the steel is modeled using  a parametrization of
a {\sc geant} simulation~\cite{geant}. It is checked with muons from \KLpimunu\ decays, in which
the pion is unambiguously identified by requiring a hadronic shower in the
CsI calorimeter~\footnote{The pion shower requirement is a CsI cluster
   with energy above 6 GeV; the probability
   for a muon to satisfy this requirement
   is less than $0.1\%$.}.  
Figure~\ref{fig:peff} shows the inefficiency of the muon
match as a function of the muon momentum.
From the distribution of distances between extrapolated
muon tracks and hit muon counters,
we conclude that
scattering in the steel is described to better
than 20\% precision.
The size of the cracks between the muon system 
counters ($\sim 1$~mm) is determined  to 50\% precision using the muon
runs.

The probability that a pion  penetrates through the  steel 
and leaves a signal in the muon system is measured in data 
using pions from  fully reconstructed  \KLpmz\ decays. This effect
is included in the MC, and is checked for \KLpimunu\ decays by studying 
the fraction of events in which  both tracks match
to hit muon system counters.
For data and MC we observe matches for both tracks in 
$2.17\%$ and $2.25\%$ of the cases, respectively.
According to the  MC, events with two matches result from pion
decays ($\sim 85\%$), pion penetrations
($\sim 10\%$) and accidental hits ($\sim 5\%$).
This study shows that the combined effect of 
pion penetrations and decays is modeled to 
better than $10\%$.

\subsubsection{Cut Variation}
We have modified various selection criteria to check the stability of 
the measured form factors.
The general strategy is to relax or tighten
one requirement at a time while 
retaining all other cuts. These requirements include the 
track separation requirement
at the drift chambers and other fiducial and kinematic cuts.

Tests leading
to significant changes
are  added to the
systematic uncertainty. For example, an extra  requirement of
no photon-like clusters reconstructed in the CsI calorimeter for the
\KLpienu\ decay mode leads to a $(-0.13\pm 0.07)\times 10^{-3}$ change
in \lplus, which is included as a systematic uncertainty.
The systematic
 uncertainties in the form factors are estimated as the sum in quadrature of 
these   variations. 

\subsubsection{Beam Shape and Kaon Energy Spectrum}
\begin{figure}
\centering
\psfig{figure=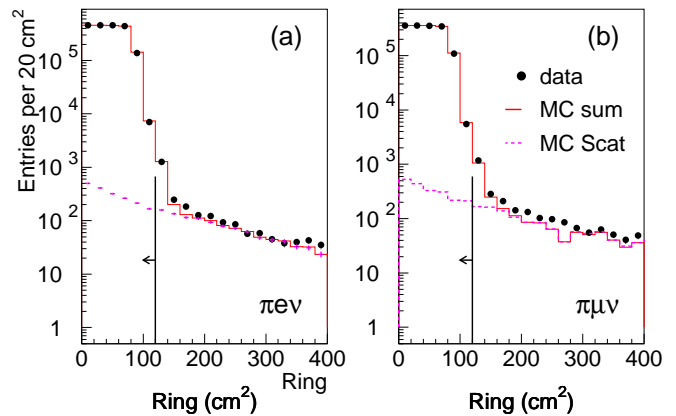,width=\linewidth}
\caption{\label{fig:ring}
Ring distributions 
for (a)  \KLpienu, and (b) \KLpimunu\
decay mode. Dots show data, ``MC sum'' refers to signal plus
background MC, and ``MC Scat'' shows the contribution from kaon scattering.
The arrow indicates the analysis requirement. Events
in the tail are mostly from kaon scattering. }
\end{figure}
The transverse 
shape of the  kaon beam is defined by a system of collimating elements,
the most important of which is  the defining 
collimator located 88~m from the primary target. 
While the geometry of the defining collimator is well
known, its exact alignment is harder to determine, leading to an
uncertainty in the beam shape near the beam edge. The beam shape 
is studied using a  ``\ring'' variable, which is calculated
using the
 X-Y vertex position projected along the kaon direction of flight to the
calorimeter surface~\cite{Alavi-Harati:2002ye}. Figure~\ref{fig:ring} compares data and MC
\ring\  distributions for the two semileptonic modes. A \ring\
value of 0~cm$^2$ 
corresponds to  the beam center and a  \ring\ value  of
$\sim 100$~cm$^2$ corresponds to the beam edge; the beam size at the
calorimeter is about $10\times10$~cm$^2$. 

In this analysis, we select events with $\ring<121$~cm$^2$.
 Larger \ring\ values
result mostly from kaon scattering in the defining collimator. 
To check if the beam edge  affects  the form factor measurement,
we repeat the analysis with  $\ring <70$~cm$^2$ and also with
$\ring <200$~cm$^2$. 
There is no statistically significant change in the form factors,
so no systematic uncertainty is assigned.

For kaon energies
between $40$ and $120$ GeV,
the kaon momentum spectrum is known to 0.01\% per GeV~\cite{ktev_kbr}.
To estimate the impact of this uncertainty on the form factor
measurement, we modify the MC energy spectrum linearly
within this uncertainty.

\subsection{Background}
\label{sec:bg}
The fraction of events in which the parent kaon has scattered in
the defining collimator, as well as
the shape of the resulting $p_{\perp,K}^2$ distribution, 
are measured with fully reconstructed
\KLpmz\ decays.
The description of  scattering is checked using events
reconstructed in the sidebands of  the ring number 
distribution~(Fig.~\ref{fig:ring});
 this comparison shows that the level of scattering is correct to better than $10\%$. 
 
\begin{figure}
\centering
\psfig{figure=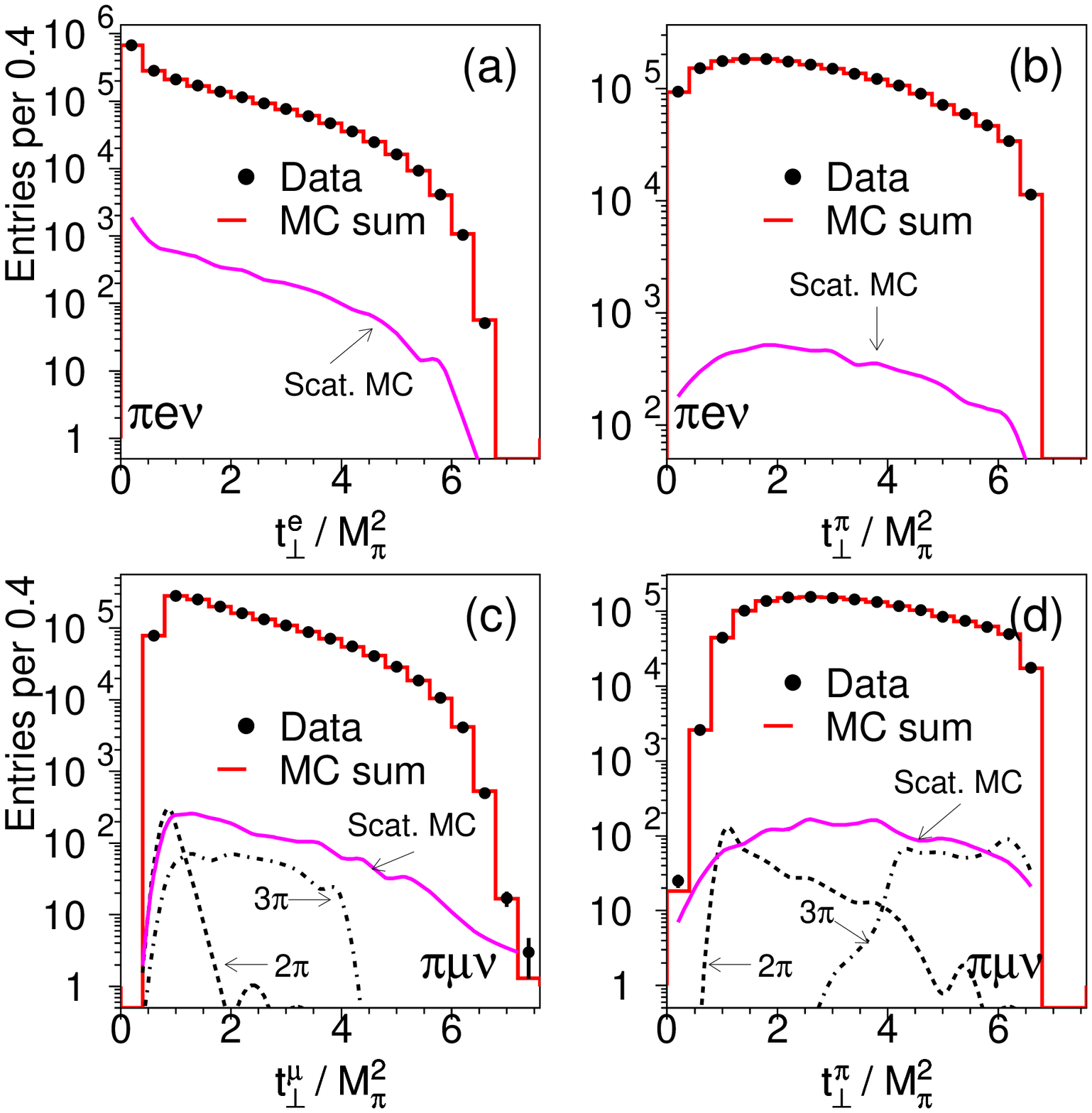,width=\linewidth}
\caption{\label{fig:tperp}
Distributions for data,
background MC and sum of background and signal MC
for (a)  $\tperpe/M^2_\pi$ in \KLpienu\ decay mode, 
(b)  $\tperpp/M^2_\pi$ in \KLpienu\ decay mode, 
(c)  $\tperpm/M^2_\pi$  in \KLpimunu\ decay mode,
and (d)  $\tperpp/M^2_\pi$ in \KLpimunu\  decay mode.}
\end{figure}
The misreconstruction background  in the \KLpimunu\ decay mode
is from  \KLpienu, \KLpmz\ and \KLpm\ decays.
This background is determined using the MC
normalized  with our branching fractions~\cite{ktev_kbr},
and is then subtracted. While 
this background is small, it is not uniform in \tperp.
Figure~\ref{fig:tperp}  shows \tperpl\ and \tperpp\
distributions for the data, signal, and  background MC.
The systematic uncertainty in the form factors is assigned as 100\%
of the change if the misreconstruction background is not subtracted
from the data.

\subsection{Fitting}
We study the systematic uncertainty arising in the fitting procedure
by comparing  \methodA\ with  \methodB\
and by varying bin sizes.
The results obtained with the  \methodA\ and \methodB\  agree with each other
within the uncorrelated statistical uncertainty. 

To evaluate the  systematic uncertainty 
caused by changes in binning, we start with the two dimensional 
\methodB\ and vary the number of bins.
The number of $M_{\pi\ell}$ bins is varied between 1 and 130, and the number
of $\tperp$ bins is varied between 1 and 80. 
Altogether, we consider about 60 fits
 between these  extremes. 
Using the deviation of each fit from the nominal fit, and
the uncorrelated statistical uncertainties,
we construct a 
$\chi^2_{bin}$, summing over all 60 fits. 
We evaluate this $\chi^2_{bin}$  for all fit
parameterizations. For example,  \lplus\ determined in
 the linear model fit to \KLpienu\ has
$\chi^2_{bin}/dof = 71.5/61$, 
and  \lzero\ determined in the linear model fit
to \KLpimunu\ has  $\chi^2_{bin}/dof = 85.9/60$.

To account for the observed non-statistical change of the
fit results with binning changes, indicated by $\chi^2_{bin}/dof>1$,
we assign an additional systematic error.
The size of this uncertainty is determined such that
addition of this error (in quadrature)
to the uncorrelated statistical uncertainty of each fit
leads to $\chi^2_{bin}/dof = 1$. 

\subsection{Radiative corrections}
\begin{figure}
\centering
\psfig{figure=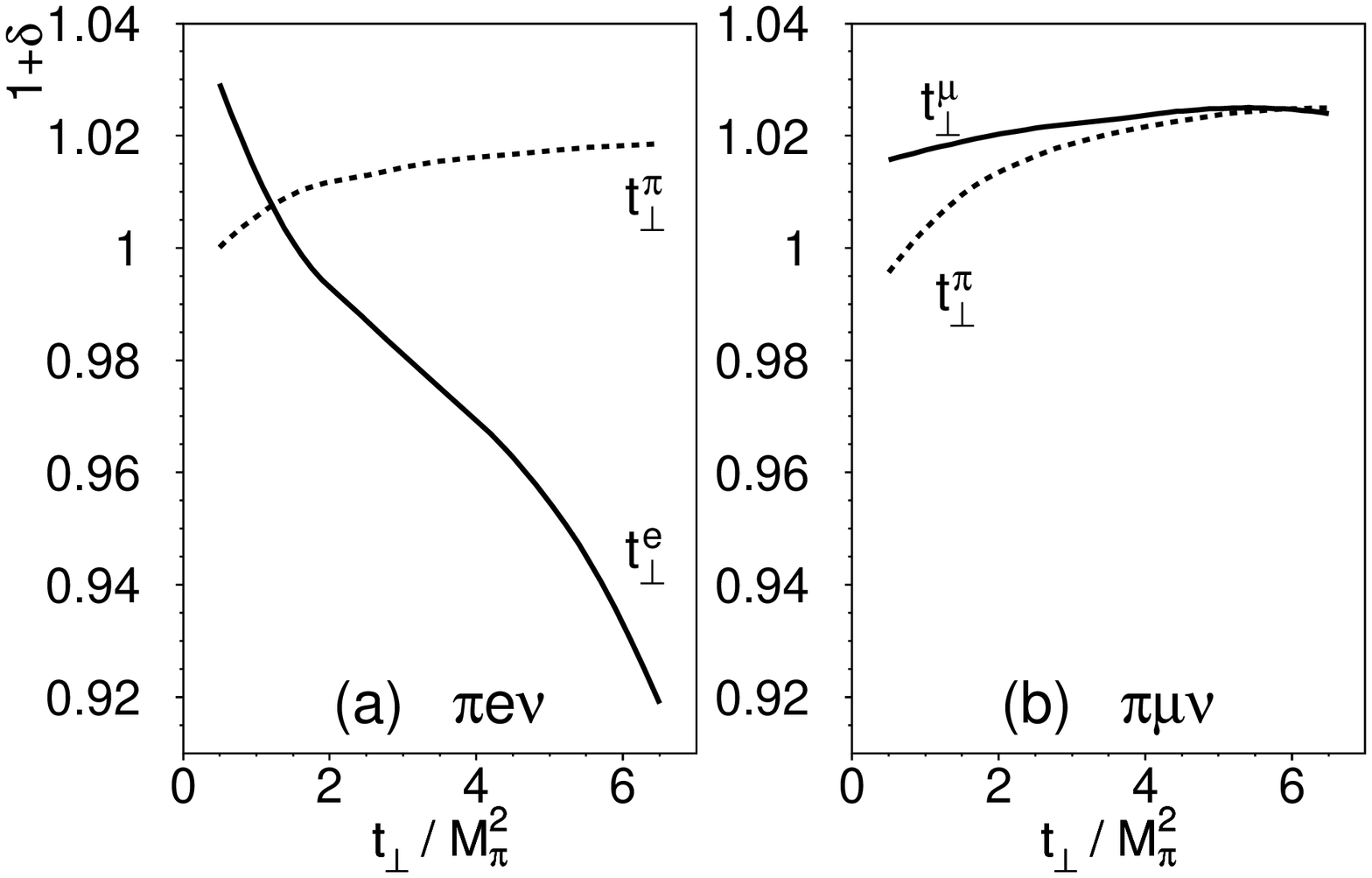,width=\linewidth}
\caption{\label{fig:radcor}
Radiative correction $1+\delta = \sigma_{rad}/\sigma_{Born}$,
calculated as a function of
(a) $\tperpe/M^2_\pi$ and $\tperpp/M^2_\pi$
for  \KLpienu,  and
(b) $\tperpm/M^2_\pi$ and $\tperpp/M^2_\pi$ for \KLpimunu. }
\end{figure}
\label{sec:radcor}
Figure~\ref{fig:radcor} shows the influence of 
radiative corrections on the  \tperpb\ distributions
for the \KLpilnu\ decay modes. 
Using the \tperpl-method, radiative corrections are 
     significantly larger for  \KLpienu\ than for
       \KLpimunu; using the \tperpp-method,
     radiative corrections are similar in both decay modes.

For the \KLpienu\ decay mode, 
variables that are based on lepton kinematics
have much
larger radiative corrections than 
  variables  based
on pion kinematics.
Consequently, the $\tperpl$-method has much greater sensitivity
to  radiative corrections than the  $\tperpp$-method.
For the $\tperpl$-method, the difference in $\lplus$
measured with the nominal MC compared to a MC without radiative
effects
 is $8\times 10^{-3}$; for the $\tperpp$-method, the corresponding difference
is only $1\times 10^{-3}$.
Using the nominal MC, the
agreement between the form factor values obtained 
using the $\tperpl$ and $\tperpp$ methods, 
$\Delta \lplus<0.5\times 10^{-3}$ at $68\%~c.l.$ 
(Sec.~\ref{sec:xc}), shows that
 radiative corrections for the \KLpienu\ decay mode
are described to better than $10\%$.

For the  \KLpimunu\ decay mode, a sensitive
check of radiative corrections is achieved by comparing
$\lzero$ measured from the
\tperpl\ and $\mpimu$ distributions in  separate 1-dimensional
fits. 
For our MC without radiative effects, the two fits
return \lzero\ values that are different
by $\Delta \lzero = (6.5\pm1.3)\times 10^{-3}$, while 
 the nominal MC gives good agreement:
$\Delta \lzero = (-0.1\pm1.3)\times 10^{-3}$. This agreement
shows that radiative corrections for the \KLpimunu\ decay mode
are described to better than $20\%$ precision.

\subsection{Summary of Systematic Uncertainties}
The total systematic uncertainty in the form factors is calculated
as the sum of systematic uncertainties in quadrature
(Table~\ref{tab:syst}). For $\lplus$ measured in
the \KLpienu\ decay mode, there are comparable contributions to the
uncertainty from
magnet kick calibration, cut variation, and
multiple scattering and resolution effects. For $\lzero$ measured in
the \KLpimunu\ decay mode, the largest systematic uncertainty is from
 modeling radiative corrections in the MC; this uncertainty is 
limited by the statistical uncertainty of the test described in Sec.~\ref{sec:radcor}.

\section{Results}
\label{sec:results}
\begin{figure}
\centering
\psfig{figure=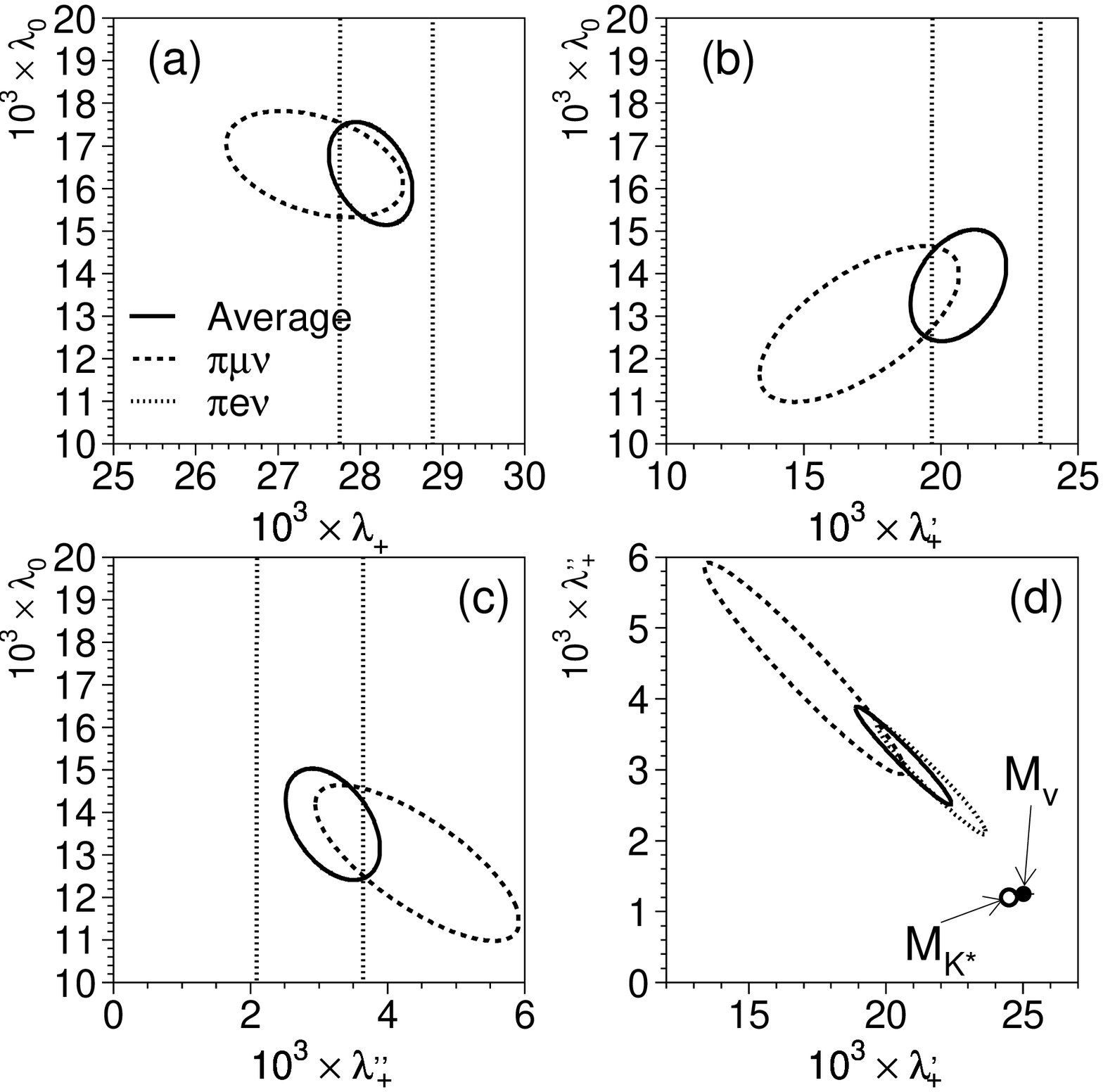,width=\linewidth}
\caption{\label{fig:results}
$1\sigma$ correlation contours  for the
(a)    linear  fit between \lzero\ and \lplus, 
(b)  quadratic fit between \lzero\ and  \lplusp, 
(c)   quadratic fit between \lzero\ and \lpluspp, and
(d)   quadratic fit between  \lpluspp\ and \lplusp. 
The dotted curve is for \KLpienu\ only, the dashed curve is
for \KLpimunu\ only, and the solid curve is for the average of the
two decay modes. 
The open and closed circle
in (d) represents \lplusp, \lpluspp\ obtained from the
Taylor expansion of the pole parametrization (Eq.~\ref{eq:taylor})
around $M_{K^*}$  and around $M_v$, respectively.  }
\end{figure}
\begin{table}
\centering
\caption{
Form factor results for both \KLpilnu\
decay modes and the average measured for Linear, Quadratic and Pole mode parametrization.
The uncertainties correspond to the total uncertainties; the breakdown 
into statistical and systematic uncertainty is given in
Table~\ref{tab:syst}. The correlation coefficients between the form
factors are given in Table~\ref{tab:corr}. $\chi^2/ndf$ for individual modes correspond to
statistical uncertainty only; $\chi^2/ndf$ for the average includes
uncorrelated systematic uncertainty. See text for explanation of
the averages.\label{tab:results}}
\begin{tabular}{c|c|c|c}
\hline
\hline
             & \KLpienu & \KLpimunu & Average \\
\hline
  \multicolumn{4}{c}{Linear Parametrization} \\
             \multicolumn{4}{c}{$(\times 10^{-3})$} \\
\hline
  \lplus    & $\lplAv \pm  \lplAt$   &  $\lplEv \pm \lplEt$ &  $ \lplav \pm \lplae$  \\
  \lzero    &                        &  $\lzEv \pm \lzEt  $ &  $ \lzav \pm \lzae $   \\
\hline
 $\chi^2/ndf$ &     \chA/\ndfA       &       \chE/\ndfE     &        \cha/1            \\
\hline
  \multicolumn{4}{c}{Quadratic Parametrization} \\
              \multicolumn{4}{c}{$(\times 10^{-3})$} \\
\hline            
  \lplusp   & $\lplaBv \pm  \lplaBt$ &  $\lplaFv \pm \lplaFt$ &  $ \lplabv \pm \lplabe$  \\
  \lpluspp  & $\lplbBv \pm  \lplbBt$ &  $\lplbFv \pm \lplbFt$ &  $ \lplbbv \pm \lplbbe$  \\
  \lzero    &                        &  $\lzFv \pm \lzFt  $ &  $ \lzbv \pm \lzbe $   \\
\hline
 $\chi^2/ndf$ &      \chB/\ndfB      &      \chF/\ndfF           &    \chb/2               \\
\hline
  \multicolumn{4}{c}{Pole Parametrization} \\
             \multicolumn{4}{c}{(MeV/c$^2$)} \\
\hline
  $M_v$     & $\lplCv \pm  \lplCt$   &  $\lplGv \pm \lplGt$ &  $ \lplcv \pm \lplce$  \\
  $M_s$     &                        &  $\lzGv \pm \lzGt  $ &  $ \lzcv \pm \lzce $   \\
\hline
 $\chi^2/ndf$ &      \chC/\ndfC      &      \chG/\ndfG      &      \chc/1             \\
\hline
\hline
\end{tabular}
\end{table}
\subsection{Form Factor results}

After all selection cuts and background subtraction, 
we have 1950543 \KLpienu\ and  1535951 \KLpimunu\
decays, respectively. Using these data, we extract \fplust\ in
the \KLpienu\ decay mode, and  both \fplust\ and \fzerot\ in the
\KLpimunu\ decay mode.
For the
nominal  \fplust\ measurement in \KLpienu, 
we use a 1-dimensional fit
to the \tperpp\ distribution
performed with \methodA. The choice of \tperpp\ over \tperpl\
is based on the much smaller sensitivity to  radiative corrections 
(see discussion in Section~\ref{sec:radcor} and Fig.~\ref{fig:radcor}).
For the nominal  \fplust\ and \fzerot\ 
measurement in  \KLpimunu, 
we use a 2-dimensional fit to the ($\mpimu,\tperpl$)
distribution performed with \methodB. The 2-dimensional fit uses the
different shape of  \fplust\ and \fzerot\  in
$\mpimu$ (Fig.~\ref{fig:lamdepl}), improving 
 the statistical precision and reducing
the correlation between the form factors compared to a 1-dimensional fit
to the $\tperp$ distribution.

The form factor results for the two decay modes and for the three different
parameterizations  
are given in  Table~\ref{tab:results};
the correlation coefficients are given in Table~\ref{tab:corr}. 
For all measurements, the
statistical
and systematic
uncertainties are comparable.
The $\Delta\chi^2 = 1$ contours
for the linear and second order fits 
based on \KLpilnu\
decays  are shown in Figure~\ref{fig:results}.
The quality of all  fits is
acceptable. On the other hand, for the  \KLpienu\ decay mode,
fits to \fplust\ using the quadratic and
pole parameterizations improve the $\chi^2$ by $18$ and 
$15$ units, respectively,
compared to the linear parametrization. Taking  the 
systematic uncertainty into account, the 
significance of the non-linear term 
in the $t$ dependence of $f_+$ is about $4\sigma$.

The KTeV result for \fplust\ is consistent with a
pole model. The fit returns a
pole mass $M_v = (\lplcv \pm \lplce)$~MeV, in fair agreement   with the
 lightest  vector $K^*$ mass 
($891.66\pm0.26$~MeV).

Lepton universality requires that the \fplust\ form factors
  be equal
for the \KLpienu\ and \KLpimunu\ decay modes. Therefore, we average 
 results for the two decay modes, taking into account statistical and systematic
correlations among the parameters, as well as correlation of the systematic
uncertainties between the decay modes. The resulting  averages are
given in the last column of Table~\ref{tab:results}. 
All parameterizations of \fplust\ are consistent for the two
decay modes. Note that although the \KLpienu\ decay mode is insensitive to
\fzerot, the correlation between \lplus\ and \lzero\ (see
  Fig.~\ref{fig:results}) results in 
an average  $\lzero$ that is different from 
 $\lzero$ deterimined only from \KLpimunu.

\begin{table*}
\centering
\caption{\label{tab:corr} 
a) Total correlation coefficients for 
      \KLpienu\  using quadratic parametrization.
     For \KLpimunu, correlation coefficients are shown for
     b) linear, c) quadratic, and d) pole parametrizations.
     For the average of the two \KLpilnu\ decay modes,
     correlation coefficients are shown for
     e) linear, f) quadratic, and g) pole parametrizations.
}
\begin{tabular}{ccccrcc}
  \begin{tabular}{c|c}
        {\bf a)}         & $ \lambda'_+$ \\
 \hline
    $\lambda''_+$ & \corB      \\
  \end{tabular}
&              ~~~
  \begin{tabular}{c|c}
        {\bf b)}        & $\lambda_0$ \\
 \hline
    $\lambda_+$ & \corE      \\
  \end{tabular}
&              ~~~
  \begin{tabular}{c|cc}
        {\bf c)}       & $\lambda_0$ & $\lambda'_+$ \\
 \hline
   $ \lambda'_+$  &  \corbF    &     \\
   $ \lambda''_+$ &  \corcF    &  \coraF \\
  \end{tabular} ~~~
& 
 \begin{tabular}{c|c}
        {\bf d)}       & $M_{s}$ \\
 \hline
   $ M_{v}$ &  \corG     \\
  \end{tabular}

&~~
 \begin{tabular}{c|c}
        {\bf e)}        & $\lambda_0$ \\
 \hline
    $\lambda_+$ & \cora     \\
  \end{tabular}
&              ~~~~~
  \begin{tabular}{c|cc}
        {\bf f)}       & $\lambda_0$ & $\lambda'_+$ \\
 \hline
   $ \lambda'_+$  &  \corbb    &     \\
   $ \lambda''_+$ &  \corcb    &  \corab \\
  \end{tabular} ~~~~~
& 
 \begin{tabular}{c|c}
        {\bf g)}       & $M_s$ \\
 \hline
   $ M_v$ &  \corc    \\
  \end{tabular}\\
\end{tabular}
\end{table*}

\subsection{Determination of Phase Space Integrals}
\label{sec:inte}
\begin{table}
\centering
\caption{Phase space integrals\label{tab:int}
for \KLpilnu\ decays, based on results in Tables~\ref{tab:results} and \ref{tab:corr}.}
\begin{tabular}{ccc}
               &  $I^e_K$           & $I^\mu_K$ \\
\hline
\hline
Linear model   &  $\ikeav \pm \ikeae$ & $\ikmav \pm \ikmae$ \\
Quadratic model& $\ikebv \pm \ikebe$  & $\ikmbv \pm \ikmbe$ \\
Pole model     &$\ikecv \pm \ikece $  & $\ikmcv \pm \ikmce$ \\
\hline
\hline
\end{tabular}
\end{table}
Using the KTeV average values of the form factors, we calculate
the decay phase space integrals (Eq.~\ref{eq:integral}).
The results obtained for the three different parameterizations are
given in Table~\ref{tab:int}. 

The quadratic parametrization
for \fplust\ leads to  integrals that are about $1\%$ lower than 
those for 
the linear parametrization. For our extraction of $|\vus|$~\cite{ktev_prl}, 
we use the
quadratic parametrization since the second order term is observed with
about
$4\sigma$ significance.The resulting decay phase space integrals are
\begin{equation}
\begin{array}{l}
I^e_K =\ikebv \pm \ikebe \pm \ikebepar \\
I^\mu_K =\ikmbv \pm \ikmbe \pm \ikmbepar,
\end{array} 
\end{equation}
where the second error is an additional systematic uncertainty based on
the difference between the quadratic and 
pole models.

\subsection{Crosschecks}
\label{sec:xc}
We have performed a series of crosschecks on the form factor measurements. 
A similar set of crosschecks is performed for the linear, quadratic and 
pole 
parameterization. Among these tests,
the best statistical sensitivity is achieved 
for the linear fit to \lplus\ for \KLpienu, and for the linear fit
to \lzero\ with fixed  \lplus\ for \KLpimunu. For  
 simplicity, we report 
crosschecks using these  linear parametrizations.

\subsubsection{Consistency among Methods}

As discussed earlier, we determine  form factors using both 
the \tperpp- and \tperpl-methods.
For $\lplus$ measured in the \KLpienu\ decay mode, 
the difference between the nominal \tperpp-method and the \tperpl-method
is $\Delta\lplus = (+0.4 \pm 0.3_{\mbox{stat}})\times 10^{-3}$, where 
the error is the uncorrelated
statistical uncertainty  estimated using an ensemble of 
MC samples. Recall from Sec.~\ref{sec:radcor} that without
radiative effects in the MC, the $\tperpp$ and
$\tperpl$ methods disagree by about $20\sigma$. 
For $\lzero$ measured in the \KLpimunu\ decay mode,
the difference between the nominal \tperpl-method and the \tperpp-method,
 $\Delta\lzero = (+0.3 \pm 0.5_{\mbox{stat}})\times 10^{-3}$, is also consistent with zero.
Similar agreement is observed for the quadratic and pole parameterizations.

Other crosschecks include one- and two-dimensional fits as
well as different form factor fitting techniques (\methodA\ and
\methodB)  explained
in Section~\ref{sec:fit}. These crosschecks do not
 show any systematic biases.

We have checked the \KLpienu\ result using an independent data
sample collected with $\times 10$ higher beam intensity.
The difference between this analysis and the nominal one
is $\Delta \lplus = (-0.3\pm 0.5_{\mbox{stat}})\times 10^{-3}$. 

The muon identification in the  \KLpimunu\ decay mode is checked with 
an  analysis in which we do not use the muon
system to identify the muon,
but instead  identify the pion by requiring a hadronic shower in the
CsI, which occurs for about $60\%$ of events.
The difference between this analysis (which 
identifies the pion) and the nominal analysis 
(which identifies the muon) is
$\Delta \lzero = (-2.4 \pm 1.5)\times 10^{-3}$. Here, the error is the
uncorrelated statistical uncertainty 
($0.8\times 10^{-3}$), combined with  the additional systematic 
uncertainty arising from the momentum dependent efficiency
 of the pion identification requirement ($1.3\times 10^{-3}$).

\subsubsection{Consistency among Data Subsets}
\label{sec:xcsub}

The stability of the results is tested  by dividing the data into
subsamples under a variety of criteria.
These include decay vertex,
minimum track momentum, track separation at the CsI calorimeter,
 pion-lepton mass, neutrino direction, magnet polarity and lepton charge.
All of these checks agree within the uncorrelated statistical uncertainty. 
A few particularly interesting  checks are discussed below.

To verify that the \tperpb-methods are not biased by the
ambiguity in the kaon energy,
we divide the data into two nearly equal subsamples based on the direction
of the neutrino in the kaon rest frame:
\begin{equation}
|\cos \theta^*_\nu| =
\sqrt{\frac{\textstyle P^*_{\parallel,\nu}\-^2}{\textstyle(P^*_{\parallel,\nu}\-^2 + P^2_{\perp,\nu})}}.
\end{equation}
 For $\cos(\theta^*_\nu)=0$, the neutrino is emitted perpendicular to the
 kaon flight direction, and there is only one kaon energy solution;
 for $|\cos(\theta^*_\nu)|=1$, the two kaon energy solutions have the maximum
 difference.  The two subsamples for this test are selected by
 $|\cos(\theta^*_\nu)|<0.4$, for which the two kaon energy solutions differ 
 on average by 18\%,
 and by  $|\cos(\theta^*_\nu)|>0.4$, 
 for which the two kaon energy solutions differ
 by 40\%.  For these two \KLpienu\ samples, we measure 
$\lplus = (28.0 \pm 0.5_{\mbox{stat}})\times 10^{-3}$
and 
$\lplus = (28.9 \pm 0.7_{\mbox{stat}})\times 10^{-3}$,
respectively,
showing  good agreement.
For \KLpimunu, there is also good agreement: 
$\lzero({|\cos\theta^*_\nu|<0.4}) -\lzero({|\cos\theta^*_\nu|>0.4}) =
(1.4 \pm 1.8_{\mbox{stat}})\times 10^{-3}$.

We check the horizontal alignment of the spectrometer
by dividing the \KLpienu\  data into four  subsamples based on 
the lepton charge and  the analyzing magnet polarity. This procedure 
separates
the data into classes 
with tracks bending in different directions in the horizontal
plane. 
We find that
for the nominal alignment, the four measurements of \lplus\ are consistent
($\chi^2/ndf = 2.95/3$); if the first drift chamber (DC1) is shifted
by $50~\mu$m in $x$-direction,
the agreement is significantly worse ($\chi^2/ndf = 30.90/3$). 
Good agreement is also observed for the same test performed for  $\lzero$
measured in the \KLpimunu\ decay mode: $\chi^2/ndf=1.8/3$.

\subsubsection{Lepton Universality}

Lepton  universality implies that the coupling constant $G_F$ and
the short-distance radiative correction $S_{EW}$ are the same 
for \KLpienu\ and \KLpimunu. Taking the ratio of Eq.~\ref{eq:vus} for these
two modes, we obtain a prediction for the partial width ratio:
\begin{equation}
 \left[ \Gpimunu/\Gpienu\right]_{\mbox{pred}}
= \frac{\textstyle 1+\deltam  }
       {\textstyle 1 + \deltae}
\cdot
\frac{\textstyle I_K^\mu}{\textstyle I_K^e}~, \label{eq:universal}
\end{equation} 
where the ratio of  radiative corrections is calculated to be
$(1+\deltam)/(1+\deltae)=\DeltaRat \pm \DeltaRatE$~\cite{Troy}.
The ratios of the phase space integrals calculated for the linear,
quadratic and pole models are:
\begin{equation}
 \frac{\textstyle I_K^{\mu}}{\textstyle I_K^{e}} =
\left\{
\begin{array}{lr}
\ikrav \pm \ikrae & \mbox{Linear  parametrization} \\
\ikrbv \pm \ikrbe & \mbox{Quadratic  parametrization} \\
\ikrcv \pm \ikrce & \mbox{Pole  parametrization}~. \\
\end{array}
\right.
\end{equation} 
Following the prescription of Sec~\ref{sec:inte},
we use the quadratic parametrization, and include
and additional error based on the pole parametrization  to
obtain $I_K^{\mu}/I_K^{e}=\ikrbv \pm \ikrbetot$.
Using
this ratio of the integrals, 
and the KTeV value of
 \Gpimunu/\Gpienu=$\RPMNvalue \pm \RPMNerrstat \pm
 \RPMNerrsyst$~\cite{ktev_kbr}, 
 we obtain
\begin{equation}
 \frac{\textstyle \Gpimunu/\Gpienu}{\textstyle [\Gpimunu/\Gpienu]_{{\mbox{pred}}}}
 =\LeptUni \pm \LeptUniE~,
\end{equation}
which shows that the KTeV 
form factors are consistent with the KTeV partial width ratio measurement.

\subsection{Comparison with other Form Factor Measurements}
\begin{figure}
\centering
\psfig{figure=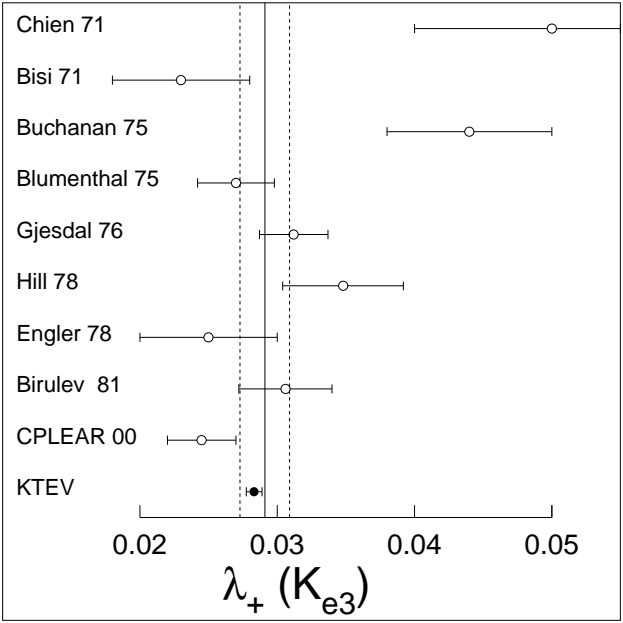,width=\linewidth}
\caption{\label{fig:lamke3}\lplus\ measured in \KLpienu\
decay mode by KTeV and by experiments used in the PDG evaluation 
(Chien 71 \cite{aspk71},
 Bisi 71 \cite{aspk71b},
 Buchanan 75 \cite{spec75},
 Blumenthal 75 \cite{spec75b},
 Gjesdal 76 \cite{spec76},
 Hill 78 \cite{strc78},
 Engler 78 \cite{hbc78},
 Birulev 81 \cite{spec81},
 CPLEAR 00 \cite{cplr00}). The PDG fit (excluding
KTeV) is shown by the vertical lines.
 }
\end{figure}
\begin{figure}
\centering
\psfig{figure=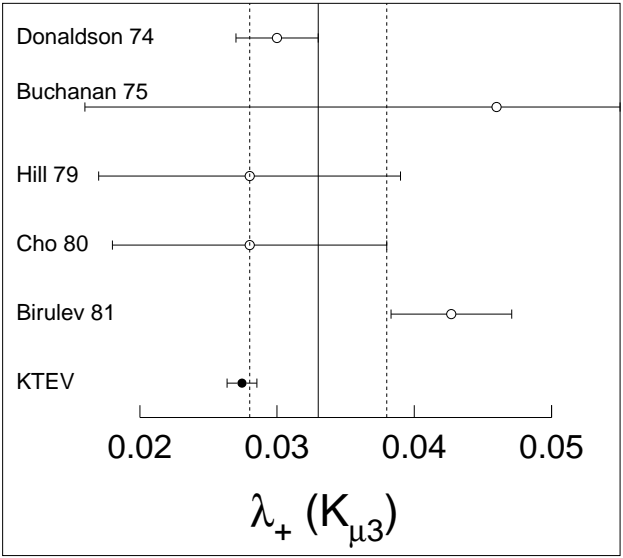,width=\linewidth}
\caption{\label{fig:lamplkm3}\lplus\ measured in \KLpimunu\
decay mode by  KTeV and by experiments  used in the PDG evaluation 
(Donaldson 74 \cite{spec74c}, 
 Buchanan 75 \cite{spec75},
 Hill 79 \cite{strc79},
 Cho  80 \cite{hbc80},
 Birulev 81 \cite{spec81}). 
The PDG fit (excluding
KTeV) is shown by the vertical lines.
 }
\end{figure}
\begin{figure}
\centering
\psfig{figure=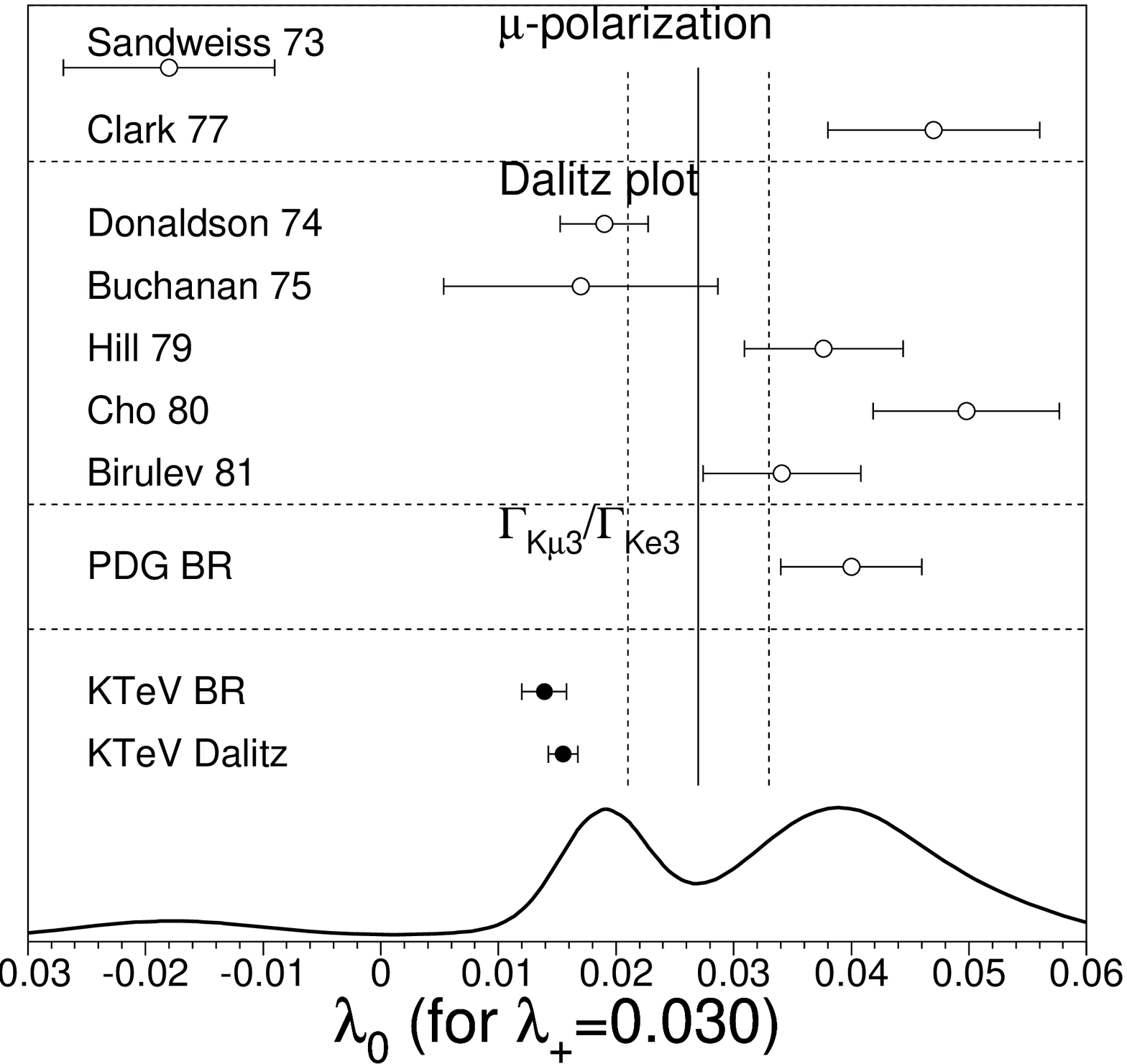,width=\linewidth}
\caption{\label{fig:lam0}\lzero\ measured  by  
 KTeV and by experiments  used in the PDG evaluation 
(Sandweiss 73 \cite{cntr73},
 Clark 77 \cite{spec77},
 Donaldson 74 \cite{spec74c},
 Buchanan 75 \cite{spec75},
 Hill 79 \cite{strc79},
 Cho 80  \cite{hbc80},
 Birulev 81 \cite{spec81}). 
PDG~BR and KTeV~BR are determined using \Gpimunu/\Gpienu\
and lepton universality. KTeV~Dalitz refers to
the technique  described in this paper.
For comparison, all central values and
uncertainties are adjusted
to 
$\lplus=0.030$,  using reported dependencies.
The PDG fit, based on $\mu$-polarization and Dalitz plot
measurements (excluding KTeV result) is shown by the vertical
lines. 
The curve represents a PDG-style ideogram constructed for all
but the two KTeV results.}
\end{figure}
Previous experiments analyzed their data using a linear
approximation for \fplust\ and \fzerot. To compare with these
results, we also use the linear parametrization.
The comparison of results for \lplus\ measured in the \KLpienu\
decay mode is shown in Fig.~\ref{fig:lamke3},
the \lplus\ comparison 
in   \KLpimunu\ is shown in Fig.~\ref{fig:lamplkm3},
and the \lzero\ comparison 
in \KLpimunu\ is shown in Fig.~\ref{fig:lam0}.
To simplify the comparison of \lzero, all measurements are adjusted
to the same  value of 
$\lplus=0.030$
using the reported dependences
~\footnote{Following the PDG prescription, the measurement
reported by Birulev~81~\cite{spec81} has not been adjusted to a different
\lplus\ value.}.

The  KTeV $\lplus$ result for \KLpienu\ is three times 
more precise than the PDG evaluation;  for \KLpimunu, our 
result is almost five times more precise.
The new KTeV measurements of \lplus\ for both
semileptonic decay modes are in agreement with most  previous
experiments and with the PDG average.
 The only significant disagreement
is  between the  measurements of  \lplus\ for   \KLpimunu\ 
performed by  KTeV and  Birulev~81~\cite{spec81}; the difference 
between these two measurements
is more than $3\sigma$ (Fig~\ref{fig:lamplkm3}).

For \lzero, the experimental situation is less clear (Fig.~\ref{fig:lam0}).
The KTeV result, which  is nearly five times more precise than
the PDG evaluation, agrees with
      only two of the eight previous measurements. 
Figure~\ref{fig:lam0} also shows  a PDG-style
ideogram  calculated without the KTeV result.
This ideogram  indicates clustering of the  measurements into two
groups, one around $\lzero \sim 0.020$
( ``middle \lzero\ group'') and another 
around $\lzero \sim 0.040$ (``high \lzero\ group''). 

Three different methods have been used to extract \lzero:
muon polarization, 
Dalitz plot density, and   \Gpimunu/\Gpienu\ 
partial width ratio  together
with the lepton universality constraint (Eq.~\ref{eq:universal}).
The largest inconsistency is a $5\sigma$ discrepancy between  the  two
 measurements based on  muon polarization (Sandweiss 73 \cite{cntr73}
and Clark 77 \cite{spec77}). 
 Sandweiss 73 is the only result with a negative \lzero.
The KTeV result does not agree with either of these  measurements.
 
Among the $\lzero$ measurements based on the Dalitz plot analysis, KTeV agrees
well with the 
highest precision measurement of
Donaldson 74 \cite{spec74c}, 
but disagrees  with the more recent
measurements (Hill~79~\cite{strc79}, Cho~80~\cite{hbc80},
 Birulev~81~\cite{spec81}).
The largest
inconsistency is a 
$4\sigma$ discrepancy  between KTeV and Cho~80~\cite{hbc80}.

The evaluation of \lzero\ from the KTeV measurement of 
\Gpimunu/\Gpienu\ agrees with the
middle \lzero\ group, while the same evaluation based on  
the PDG fit lands in the high \lzero\ group. 
The PDG value for \Gpimunu/\Gpienu\
 is to a large extent driven by the  measurement 
reported by  Cho 80 \cite{hbc80}, which disagrees with 
the analogous KTeV measurement by more than  $3\sigma$.

It is worth mentioning that the physics interpretations 
of the middle and high \lzero\ groups 
are quite different. 
For the former, the
form factors are consistent with a pole  model
($\lambda_0 < \lambda_+$), and also agree  
with those measured  for $K^\pm$~\footnote{The $K^\pm$ form factors from
PDG are 
$\lplus(K^+_{e3}) = 0.0283\pm0.0015$, $\lplus(K^+_{\mu3})=0.033\pm 0.010$,
and $\lzero = 0.013\pm 0.005$.
Recently, the ISTRA+ collaboration reported
new measurements of $\lplus(K^+_{\mu3})=0.0277\pm 0.0016$  
and $\lzero = 0.0183\pm 0.0013$~\cite{istra}.}, suggesting that isospin symmetry
breaking effects are small. 
For the high \lzero\ group, where $\lambda_0  > \lambda_+$ ($2\sigma$),
the  simple pole model is disfavoured; 
the high \lzero\ group also shows   
more than $3\sigma$ significance of  isospin symmetry breaking
for \fzerot.

\section{Conclusions}
In this paper, we have presented  new measurements of the
form factors in  semileptonic $K_L$ decays performed 
for \KLpienu\ and \KLpimunu.
The measurement of \fplust\ is consistent
for the two decay modes, and provides about a threefold
increase in precision compared to the PDG average. 
KTeV measurements of \fplust\ are in good agreement
with those used in the PDG evaluation.

The KTeV result for \lzero,
which is nearly five times
         more precise than the PDG average, disagrees with
         six of the eight measurements used in the PDG average.
         Despite the poor agreement between KTeV and previous results, 
         the most precise measurement~\cite{spec74c} used in the PDG 
         evaluation agrees  with the KTeV measurement.

The new KTeV result indicates 
 the 
presence of  a non-linear term in the $t$ dependence
of $f_+$ with about $4\sigma$ significance.
This non-linear term  is consistent with  pole model expectations.
The KTeV result for  \lzero\ is also consistent with the pole model
prediction, and  is in agreement with  the value measured in $K^\pm$ decays,
 suggesting that isospin breaking effects are small.
   
Using the KTeV measurements of the semileptonic form factors,
we have calculated the values of the  
phase space integrals for the \KLpienu\ and \KLpimunu\ decay modes. 
We find that inclusion of the non-linear term in the  \fplust\ expansion 
reduces  the value of these integrals by about $1\%$.  
Lepton universality holds  for the KTeV data with $0.5\%$ precision, showing
consistency between the KTeV form factor measurements presented in this paper,
KTeV semileptonic  partial width ratios
reported in~\cite{ktev_kbr}, and  radiative corrections calculated
in~\cite{Troy}.

\section{Acknowledgments}

We gratefully acknowledge the support and effort of the Fermilab
staff and the technical staffs of the participating institutions for
their vital contributions.  This work was supported in part by the U.S. 
Department of Energy, The National Science Foundation, and The Ministry of
Education and Science of Japan.

%

\begin{thebibliography}{31}
\expandafter\ifx\csname natexlab\endcsname\relax\def\natexlab#1{#1}\fi
\expandafter\ifx\csname bibnamefont\endcsname\relax
  \def\bibnamefont#1{#1}\fi
\expandafter\ifx\csname bibfnamefont\endcsname\relax
  \def\bibfnamefont#1{#1}\fi
\expandafter\ifx\csname citenamefont\endcsname\relax
  \def\citenamefont#1{#1}\fi
\expandafter\ifx\csname url\endcsname\relax
  \def\url#1{\texttt{#1}}\fi
\expandafter\ifx\csname urlprefix\endcsname\relax\def\urlprefix{URL }\fi
\providecommand{\bibinfo}[2]{#2}
\providecommand{\eprint}[2][]{\url{#2}}

\bibitem[{\citenamefont{Cabibbo}(1963)}]{cabibbo}
\bibinfo{author}{\bibfnamefont{N.}~\bibnamefont{Cabibbo}},
  \bibinfo{journal}{Phys.\ Rev.\ Lett.} \textbf{\bibinfo{volume}{10}},
  \bibinfo{pages}{531} (\bibinfo{year}{1963}).

\bibitem[{\citenamefont{Kobayashi and Maskawa}(1973)}]{km}
\bibinfo{author}{\bibfnamefont{M.}~\bibnamefont{Kobayashi}} \bibnamefont{and}
  \bibinfo{author}{\bibfnamefont{T.}~\bibnamefont{Maskawa}},
  \bibinfo{journal}{Prog.\ Theor.\ Phys.} \textbf{\bibinfo{volume}{49}},
  \bibinfo{pages}{652} (\bibinfo{year}{1973}).

\bibitem[{\citenamefont{Leutwyler and Roos}(1984)}]{leut-roos}
\bibinfo{author}{\bibfnamefont{H.}~\bibnamefont{Leutwyler}} \bibnamefont{and}
  \bibinfo{author}{\bibfnamefont{M.}~\bibnamefont{Roos}}, \bibinfo{journal}{Z.\
  Phys.} \textbf{\bibinfo{volume}{C25}}, \bibinfo{pages}{91}
  (\bibinfo{year}{1984}).

\bibitem[{\citenamefont{Alexopoulos et~al.}(2004{\natexlab{a}})}]{ktev_prl}
\bibinfo{author}{\bibfnamefont{T.}~\bibnamefont{Alexopoulos}}
  \bibnamefont{et~al.} (\bibinfo{collaboration}{KTeV})
  (\bibinfo{year}{2004}{\natexlab{a}}), \bibinfo{note}{submitted to Phys. Rev.
  Lett.}, \eprint{hep-ex/0406001}.

\bibitem[{\citenamefont{{Particle Data Group}}(2002)}]{pdg02}
\bibinfo{author}{\bibnamefont{{Particle Data Group}}}, \bibinfo{journal}{Phys.\
  Ref.\ D} \textbf{\bibinfo{volume}{66}}, \bibinfo{pages}{1}
  (\bibinfo{year}{2002}).

\bibitem[{\citenamefont{Apostolakis et~al.}(2000)}]{cplr00}
\bibinfo{author}{\bibfnamefont{A.}~\bibnamefont{Apostolakis}}
  \bibnamefont{et~al.}, \bibinfo{journal}{Phys.\ Lett.}
  \textbf{\bibinfo{volume}{B473}}, \bibinfo{pages}{186} (\bibinfo{year}{2000}).

\bibitem[{\citenamefont{{Particle Data Group}}(1982)}]{pdg82}
\bibinfo{author}{\bibnamefont{{Particle Data Group}}}, \bibinfo{journal}{Phys.\
  Lett.} \textbf{\bibinfo{volume}{111B}}, \bibinfo{pages}{{1}}
  (\bibinfo{year}{1982}).

\bibitem[{\citenamefont{Battle et~al.}(1994)}]{cleo_tau}
\bibinfo{author}{\bibfnamefont{H.}~\bibnamefont{Battle}} \bibnamefont{et~al.},
  \bibinfo{journal}{Phys.\ Rev.\ Lett.} \textbf{\bibinfo{volume}{73}},
  \bibinfo{pages}{1079} (\bibinfo{year}{1994}).

\bibitem[{\citenamefont{Abbiendi et~al.}(2000)}]{tau_opal}
\bibinfo{author}{\bibfnamefont{G.}~\bibnamefont{Abbiendi}}
  \bibnamefont{et~al.}, \bibinfo{journal}{Eur.\ Phys.\ J.\ C}
  \textbf{\bibinfo{volume}{13}}, \bibinfo{pages}{213} (\bibinfo{year}{2000}).

\bibitem[{\citenamefont{Barate et~al.}(1999)}]{tau_aleph}
\bibinfo{author}{\bibfnamefont{R.}~\bibnamefont{Barate}} \bibnamefont{et~al.},
  \bibinfo{journal}{Eur.\ Phys.\ J.\ C} \textbf{\bibinfo{volume}{10}},
  \bibinfo{pages}{1} (\bibinfo{year}{1999}).

\bibitem[{\citenamefont{Marciano and Sirlin}(1986)}]{sew}
\bibinfo{author}{\bibfnamefont{W.}~\bibnamefont{Marciano}} \bibnamefont{and}
  \bibinfo{author}{\bibfnamefont{A.}~\bibnamefont{Sirlin}},
  \bibinfo{journal}{Phys.\ Rev.\ Lett.} \textbf{\bibinfo{volume}{56}},
  \bibinfo{pages}{22} (\bibinfo{year}{1986}).

\bibitem[{\citenamefont{Ginsberg}(1966)}]{Ginsberg}
\bibinfo{author}{\bibfnamefont{E.}~\bibnamefont{Ginsberg}},
  \bibinfo{journal}{Phys.\ Rev.} \textbf{\bibinfo{volume}{142}},
  \bibinfo{pages}{1035} (\bibinfo{year}{1966}).

\bibitem[{\citenamefont{Andre}(2004)}]{Troy}
\bibinfo{author}{\bibfnamefont{T.}~\bibnamefont{Andre}} (\bibinfo{year}{2004}),
  \bibinfo{note}{to be submitted to Phys. Rev. D}, \eprint{hep-ph/0406006}.

\bibitem[{\citenamefont{Brandenburg et~al.}(1973)}]{Brandenburg73}
\bibinfo{author}{\bibfnamefont{G.}~\bibnamefont{Brandenburg}}
  \bibnamefont{et~al.}, \bibinfo{journal}{Phys.\ Rev.}
  \textbf{\bibinfo{volume}{D8}}, \bibinfo{pages}{1978} (\bibinfo{year}{1973}).

\bibitem[{\citenamefont{Alavi-Harati et~al.}(2003)}]{Alavi-Harati:2002ye}
\bibinfo{author}{\bibfnamefont{A.}~\bibnamefont{Alavi-Harati}}
  \bibnamefont{et~al.} (\bibinfo{collaboration}{KTeV}), \bibinfo{journal}{Phys.
  Rev.} \textbf{\bibinfo{volume}{D67}}, \bibinfo{pages}{012005}
  (\bibinfo{year}{2003}), \eprint{hep-ex/0208007}.

\bibitem[{\citenamefont{Alexopoulos et~al.}(2004{\natexlab{b}})}]{ktev_kbr}
\bibinfo{author}{\bibfnamefont{T.}~\bibnamefont{Alexopoulos}}
  \bibnamefont{et~al.} (\bibinfo{collaboration}{KTeV})
  (\bibinfo{year}{2004}{\natexlab{b}}), \bibinfo{note}{submitted to Phys. Rev.
  D}, \eprint{hep-ex/0406002}.

\bibitem[{\citenamefont{Brun et~al.}(1994)}]{geant}
\bibinfo{author}{\bibfnamefont{R.}~\bibnamefont{Brun}} \bibnamefont{et~al.}
  (\bibinfo{year}{1994}), \bibinfo{note}{{\sc geant} 3.21, CERN, Geneva}.

\bibitem[{\citenamefont{Chien et~al.}(1971)}]{aspk71}
\bibinfo{author}{\bibfnamefont{C.}~\bibnamefont{Chien}} \bibnamefont{et~al.},
  \bibinfo{journal}{Phys.\ Lett.} \textbf{\bibinfo{volume}{35B}},
  \bibinfo{pages}{261} (\bibinfo{year}{1971}).

\bibitem[{\citenamefont{Bisi et~al.}(1971)}]{aspk71b}
\bibinfo{author}{\bibfnamefont{V.}~\bibnamefont{Bisi}} \bibnamefont{et~al.},
  \bibinfo{journal}{Phys.\ Lett.} \textbf{\bibinfo{volume}{36B}},
  \bibinfo{pages}{533} (\bibinfo{year}{1971}).

\bibitem[{\citenamefont{Buchanan et~al.}(1975)}]{spec75}
\bibinfo{author}{\bibfnamefont{C.}~\bibnamefont{Buchanan}}
  \bibnamefont{et~al.}, \bibinfo{journal}{Phys.\ Rev.}
  \textbf{\bibinfo{volume}{D11}}, \bibinfo{pages}{457} (\bibinfo{year}{1975}).

\bibitem[{\citenamefont{Blumenthal et~al.}(1975)}]{spec75b}
\bibinfo{author}{\bibfnamefont{R.}~\bibnamefont{Blumenthal}}
  \bibnamefont{et~al.}, \bibinfo{journal}{Phys.\ Rev.\ Lett.}
  \textbf{\bibinfo{volume}{34}}, \bibinfo{pages}{164} (\bibinfo{year}{1975}).

\bibitem[{\citenamefont{Gjesdal et~al.}(1976)}]{spec76}
\bibinfo{author}{\bibfnamefont{G.}~\bibnamefont{Gjesdal}} \bibnamefont{et~al.},
  \bibinfo{journal}{Nucl.\ Phys.} \textbf{\bibinfo{volume}{B109}},
  \bibinfo{pages}{118} (\bibinfo{year}{1976}).

\bibitem[{\citenamefont{Hill et~al.}(1978)}]{strc78}
\bibinfo{author}{\bibfnamefont{D.}~\bibnamefont{Hill}} \bibnamefont{et~al.},
  \bibinfo{journal}{Phys.\ Lett.} \textbf{\bibinfo{volume}{73B}},
  \bibinfo{pages}{483} (\bibinfo{year}{1978}).

\bibitem[{\citenamefont{Engler et~al.}(1978)}]{hbc78}
\bibinfo{author}{\bibfnamefont{A.}~\bibnamefont{Engler}} \bibnamefont{et~al.},
  \bibinfo{journal}{Phys.\ Rev.} \textbf{\bibinfo{volume}{D18}},
  \bibinfo{pages}{623} (\bibinfo{year}{1978}).

\bibitem[{\citenamefont{Birulev et~al.}(1981)}]{spec81}
\bibinfo{author}{\bibfnamefont{V.}~\bibnamefont{Birulev}} \bibnamefont{et~al.},
  \bibinfo{journal}{Nucl.\ Phys.} \textbf{\bibinfo{volume}{B182}},
  \bibinfo{pages}{1} (\bibinfo{year}{1981}).

\bibitem[{\citenamefont{Donaldson et~al.}(1974)}]{spec74c}
\bibinfo{author}{\bibfnamefont{G.}~\bibnamefont{Donaldson}}
  \bibnamefont{et~al.}, \bibinfo{journal}{Phys.\ Rev.}
  \textbf{\bibinfo{volume}{D9}}, \bibinfo{pages}{2960} (\bibinfo{year}{1974}).

\bibitem[{\citenamefont{Hill et~al.}(1979)}]{strc79}
\bibinfo{author}{\bibfnamefont{D.}~\bibnamefont{Hill}} \bibnamefont{et~al.},
  \bibinfo{journal}{Nucl.\ Phys.} \textbf{\bibinfo{volume}{B153}},
  \bibinfo{pages}{39} (\bibinfo{year}{1979}).

\bibitem[{\citenamefont{Cho et~al.}(1980)}]{hbc80}
\bibinfo{author}{\bibfnamefont{Y.}~\bibnamefont{Cho}} \bibnamefont{et~al.},
  \bibinfo{journal}{Phys.\ Rev.} \textbf{\bibinfo{volume}{D22}},
  \bibinfo{pages}{2688} (\bibinfo{year}{1980}).

\bibitem[{\citenamefont{Sandweiss et~al.}(1973)}]{cntr73}
\bibinfo{author}{\bibfnamefont{J.}~\bibnamefont{Sandweiss}}
  \bibnamefont{et~al.}, \bibinfo{journal}{Phys.\ Rev.\ Lett.}
  \textbf{\bibinfo{volume}{30}}, \bibinfo{pages}{1002} (\bibinfo{year}{1973}).

\bibitem[{\citenamefont{Clark et~al.}(1977)}]{spec77}
\bibinfo{author}{\bibfnamefont{A.}~\bibnamefont{Clark}} \bibnamefont{et~al.},
  \bibinfo{journal}{Phys.\ Rev.} \textbf{\bibinfo{volume}{D15}},
  \bibinfo{pages}{553} (\bibinfo{year}{1977}).

\bibitem[{\citenamefont{Yushchenko et~al.}(2004)}]{istra}
\bibinfo{author}{\bibfnamefont{O.}~\bibnamefont{Yushchenko}}
  \bibnamefont{et~al.}, \bibinfo{journal}{Phys.\ Lett.}
  \textbf{\bibinfo{volume}{B581}}, \bibinfo{pages}{31} (\bibinfo{year}{2004}).

\end{thebibliography}

\end{document}